# Designing Secure Interconnects for Modern Microelectronics: From SoCs to Emerging Chiplet-Based Architectures

Dipal Halder

Department of Electrical and Computer Engineering

University of Florida

dipal.halder@ufl.edu

**Abstract**

The globalization of semiconductor supply chains has exposed Network-on-Chip (NoC) interconnects in System-on-Chip (SoC) architectures to critical security risks, including reverse engineering and IP theft. To address these threats, this work builds on two methodologies: ObNoCs [11], which obfuscates NoC topologies using programmable multiplexers, and POTENT [10], which enhances post-synthesis security against SAT-based attacks. These techniques ensure robust protection of NoC interconnects with minimal performance overhead.

As the industry shifts to chiplet-based heterogeneous architectures, this research extends ObNoCs and POTENT to secure intra- and inter-chiplet interconnects. New challenges, such as safeguarding inter-chiplet communication and interposer design, are addressed through enhanced obfuscation, authentication, and encryption mechanisms. Experimental results demonstrate the practicality of these approaches for high-security applications, ensuring trust and reliability in monolithic and modular systems.

## 1 Introduction

The explosive growth of complex microelectronic systems has made Network-on-Chip (NoC) architectures indispensable for enabling efficient communication between integrated hardware blocks. As systems evolve from traditional System-on-Chip (SoC) designs to emerging chiplet-based architectures, the security of these interconnect fabrics has become increasingly critical. The globalization of semiconductor supply chains introduces vulnerabilities at multiple stages - from design and fabrication to assembly and testing - where malicious actors can exploit NoC topologies through reverse-engineering attacks. These security threats are particularly concerning because NoC topologies contain sensitive information about system-level parameters, including inter-IP communication patterns, workload characteristics, and operational priorities. An adversary who successfully reverse-engineers these interconnects can potentially deduce critical design features, enable unauthorized chip production, or insert hardware Trojans. The challenge becomes even more complex in chiplet-based systems, where both intra-chiplet and inter-chiplet communications must be secured against increasingly sophisticated attacks. Previous research has established foundational approaches to securing NoC architectures. System-level obfuscation techniques, such as ObNoCs, demonstrate the effectiveness of programmable multiplexers in concealing intended topologies within a space of valid configurations. POTENT extends this security framework by implementing post-synthesis obfuscation at the gate-level netlist, providing robust protection against synthesis-induced vulnerabilities and Boolean satisfiability (SAT) attacks. However, the transition to chiplet-based architectures introduces new security challenges that existing methodologies do not fully address. This research proposes a comprehensive framework for securing interconnect fabrics across both traditional SoC and emerging chiplet-based architectures. We aim to extend established obfuscation techniques while developing novel methods specifically tailored to protect inter-chiplet communication channels, addressing unique challenges such as die-to-die interfaces, interposer designs, and modular system integration. Our approach combines advanced topology obfuscation with innovative authentication and encryption mechanisms, ensuring security without compromising the performance benefits that drive the adoption of these modern architectures. Through this work, we seek to establish a new paradigm in secure interconnect design that addresses the full spectrum of contemporary microelectronic systems, from monolithic SoCs to sophisticated chiplet-based implementations, while maintaining minimal overhead and maximum design flexibility.

This work makes the following important contributions.

- **NoC Topology and Gate-Level Obfuscation:** Introduced a programmable MUX-based framework (ObNoCs) for dynamic topology obfuscation and extended it to post-synthesis gate-level obfuscation (POTENT) to address reverse-engineering threats at multiple stages of the design flow.

- **Defense Against SAT Attacks:** Enhanced resilience against Boolean satisfiability (SAT)−based de-obfuscation techniques, ensuring robust protection against adversarial analysis.

- **Scalable and Practical Security:** Achieved low overhead in terms of area, power, and latency while maintaining high levels of security, making these solutions suitable for diverse SoC designs and high-security applications.

- **Comprehensive Analysis:** Demonstrated effectiveness through rigorous experimental validation on industry-standard platforms, ensuring applicability in realistic use cases.

# 2  Background

## 2.1  NoC Fabrics

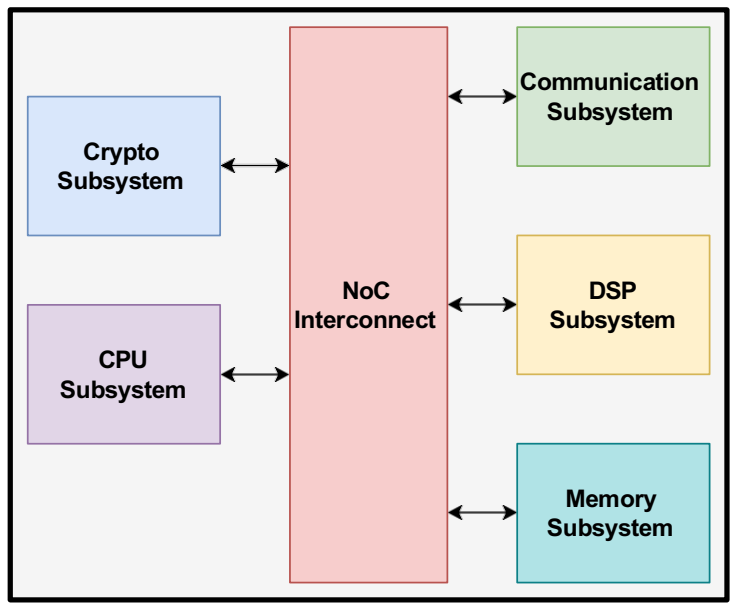

Figure 1: Modern System-on-Chip Design

The primary coordination mechanism for IPs in an SoC is message-based communication. Consequently, the communication fabrics constitute a crucial component of an SoC design. Traditional communication fabrics used point-to-point, crossbar, or bus architectures. More recently, NoC fabrics have gained popularity in industrial SoCs. An NoC involves a collection of routers connected to realize a target topology. Many industrial SoC designs make use of a tree topology, although other networks such as cycle, mesh, or torus are also in use [1]. Routers in an NoC typically include configurable routing tables which can be reconfigured if necessary by the operating system at boot-time. Fig **1** shows a model system-on-chip design which has different subsystems for various functionalities. In Fig **2**, all of the subsystems are elaborated where each subsystem consist of numerous IPs, connected together with router based communication system. A key advantage of NoC architecture is the flexibility provided in implementing power management with low overhead as it is possible to simply shut off (or reducing high-speed functionality) routers in the sub-network when message communication through the sub-network is reduced. Note of course that these benefits do come with a number of challenges, including complex optimization requirements for achieving power utilization, fault tolerance, quality of service (QoS), and CAD support for NoC design [12]. For instance, to address the challenge of high throughput performance, Pathania *et al.* [15] introduced a unique topology-based performance heterogeneity which exploits many-core heterogeneity to extract more performance. Indeed, our work also accounts for resource utilization as an overhead metric to trade off for achieving security goals.

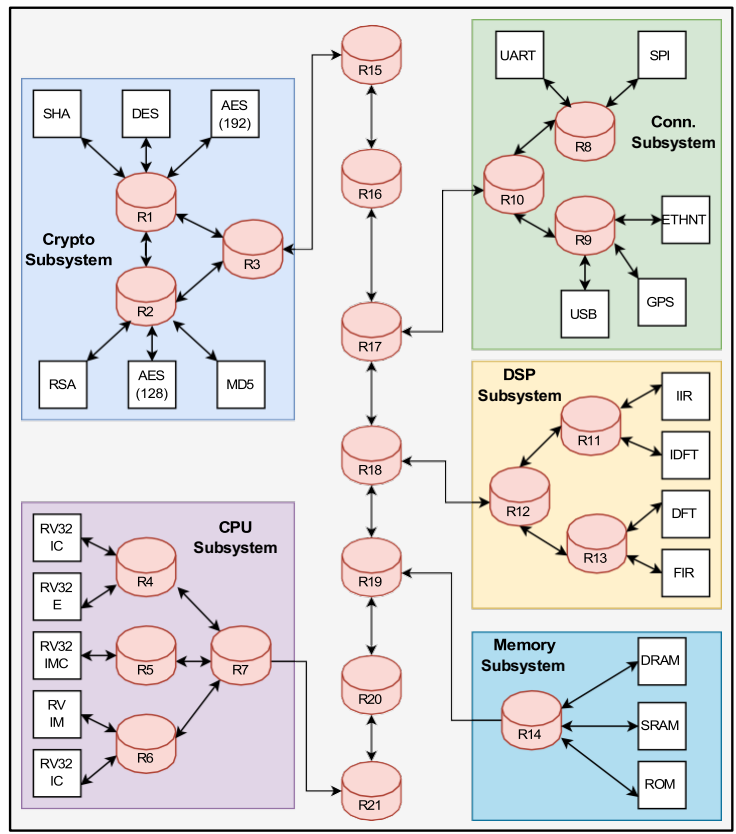

Figure 2: Expanded view of an industrial System-on-Chip Design

## 2.2   SoC Supply Chain Security

Semiconductor design has evolved over the past decade into a global enterprise incorporating 3PIP (third-party IP) vendors, IC design houses, fabrication labs, and testing facilities dispersed over multiple countries and continents. The globalization in the supply chain has been driven by a number of factors, including aggressive time-to-market requirements, miniaturization of VLSI technology, increased fabrication and validation costs, etc. In particular, the exponential shrinkage of transistor nodes over the past decades has enabled the IC designers to pack complex, multi-core and many-core designs with advanced performance in area and power-constrained chip designs, with a resultant increase in the price of fabrication. As a result, a majority of the semiconductor companies are going fabless and outsourcing chip fabrication to globally distributed remote foundries. An unfortunate upshot is that it is possible for a rogue foundry to mount a variety of attacks subverting the design and production of the system. One class of attacks involves alteration of the design by inserting malicious Trojan circuitry. Another class of attacks includes reverse engineering, possibly targeting the entire SoC or some of the key IPs. Other supply chain attacks include cloning, piracy, counterfeiting, recycling, and overproduction are shown in Fig 3.

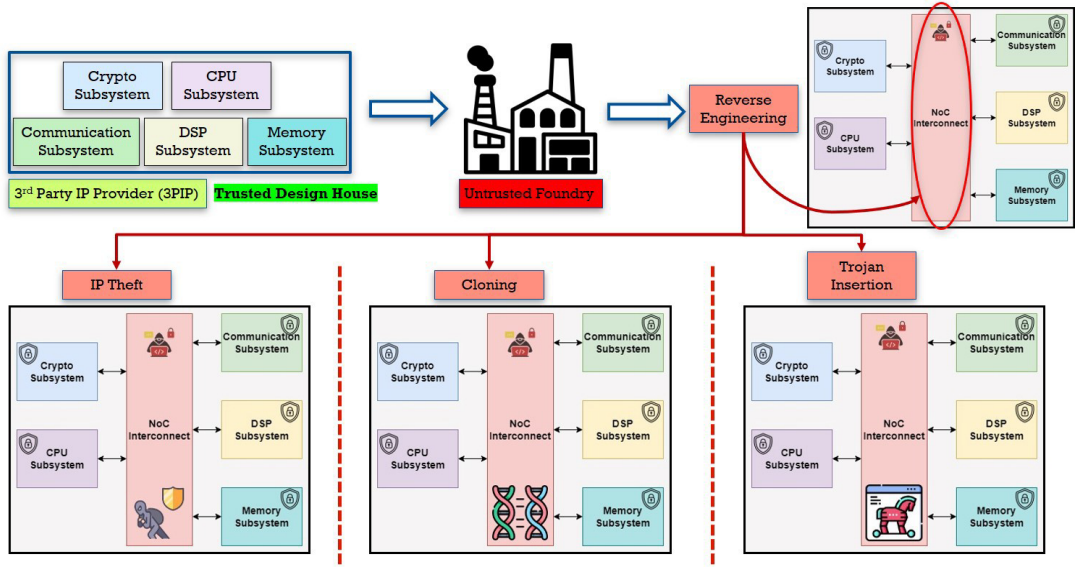

Figure 3: Different security vulnerabilities in Network-on-Chip Design

## 2.3   Reverse-Engineering Attacks

Reverse engineering is the technique of extracting data from a chip by dissecting its functional elements and constituent parts. In the context of NoCs, the goal of the reverse-engineering attacks is to extract information of the NoC topology from an SoC implementation (either as netlist or as a fabricated silicon). Although there have not been dedicated techniques specifically targeted to reverse-engineer NoC topologies, this can be accomplished by common methods such as imaging, side-channel analysis, and micro probing all of which have been extensively researched. Torrance *et al.* have discussed several types of reverse engineering, of which the products are disassembling, system-level analysis, and circuit extraction in [21]. Gomez *et al.* described a method that uses image processing to recreate a chip's gate netlist after it has been decapped and delayered. Using this method, the whole netlist is rebuilt by teaching an image processing tool to identify common cell layout patterns and extract routing information [9]. Additionally, there are purely algorithmic ways to reverse engineering digital circuits for the purpose of inferring a high-level netlist with components like register files, adders, and counters from unstructured netlists. For instance, Subramanyan *et al.* [20] segmented the netlist into potential/candidate modules and by determining functionality using methods akin to design synthesis. The goal is to make it easier for a human analyst to comprehend the functionality of an unstructured netlist by identifying as many components as possible. Holler *et al.* [13] describe a method that combines image processing and machine learning to improve the effectiveness and precision of reverse engineering procedures. Botero *et al.* have used machine learning algorithms to recognize patterns and categorize the parts, while image processing techniques can be used to extract features from photos of the hardware [7]. Recently, there has also been work on discovering the algorithms to find the specification of the extracted design [6].

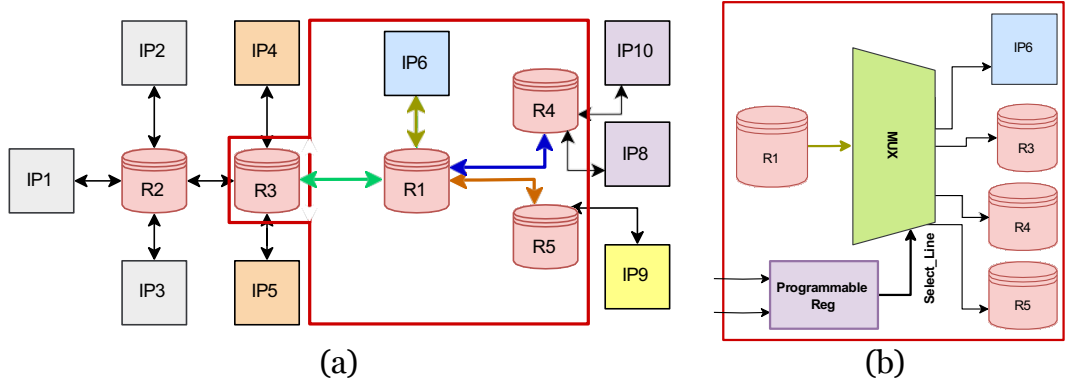

Figure 4: NoC Interconnect Transformation Topology Obfuscation. (a) An NoC-based SoC Design Example with a simple Tree Topology. (b) Transformation of the outgoing edge of router R1

# 3 Developed Framework for NoC Security

In the rapidly evolving landscape of System-on-Chip (SoC) design, securing the interconnect fabric has become a cornerstone of protecting intellectual property and ensuring system integrity. To address these challenges, a generic framework was developed to fortify the Network-on-Chip (NoC) interconnect against reverse engineering and other hardware-based attacks. This framework leverages two complementary solutions—ObNoCs and POTENT—to achieve robust obfuscation across different stages of the SoC design flow. The ObNoCs methodology introduces topology obfuscation at the architectural level, safeguarding the NoC fabric from adversaries attempting to extract sensitive communication patterns. Meanwhile, POTENT extends these protections to the post-synthesis stage, addressing obfuscation vulnerabilities introduced by synthesis tools and optimization processes. Together, these solutions provide a holistic approach to protecting NoC interconnects in SoC designs, balancing security, performance, and scalability while setting the foundation for securing future heterogeneous systems, including chiplet-based architectures.

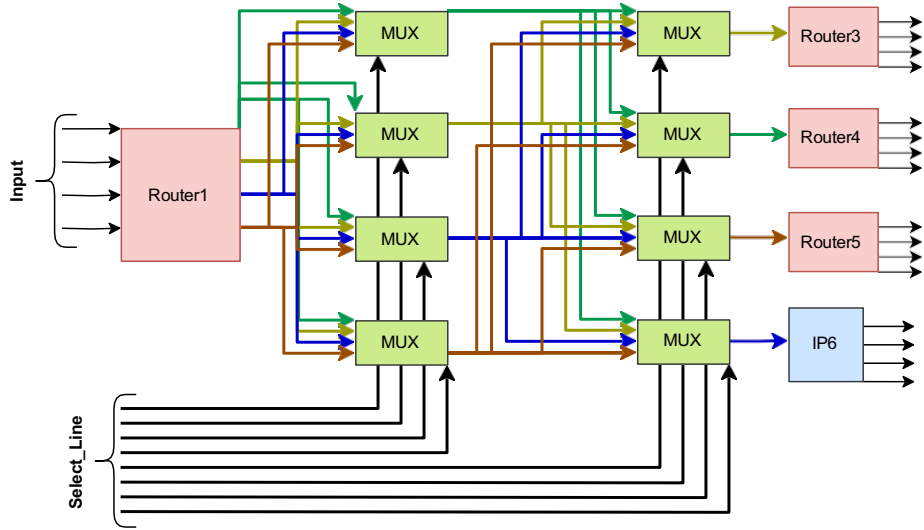

Figure 5: Example of custom MUX based transformation for outgoing signals of Router1 to IP6, Router 3, Router 4 ,and Router 5

## 3.1  ObNoCs Architecture

Consider the SoC design shown in Fig. 4(a). It includes an NoC with five routers organized in a simple tree network. Suppose we want to redact the connection $R_1 \rightarrow$ IP$_6$, Fig. 4(b) shows the method to transform this connection. In particular, $R_1$ is connected to the input of a $1 \times 4$ demultiplexer and the output of the demultiplexer is connected to the four different hardware blocks. We call this circuitry the *DEMUX switch*. The control of the DEMUX switch is connected to a register that can be programmed after fabrication, and the control bits determine which IP is actually connected to $R_1$ in this port. If the register is programmed with the bits 00 then $R_1$ would be connected in this port to IP$_6$ which corresponds to the original topology. Note that each pattern of bits from 00 through 11 corresponds to *some connection*, *e.g.*, 01 would correspond to $R_1 \rightarrow R_3$ and 10 to $R_1 \rightarrow R_4$.

Fig. 5 shows how the transformation above would be implemented for router $R_1$. In particular, we exercise $4 \times 1$ MUX to achieve a *MUX-DEMUX* based switch for each port of the router that we need to redact. This makes the architecture of ObNoCs 2-stage where the first stage acts as the DEMUX switch and the second stage as the MUX switch. The control bits for this *MUX-DEMUX* switch are connected to a register which can be programmed at runtime.

From the example, it is clear that the NoC obfuscated by ObNoCs would have the same topological behavior as the unobfuscated NoC if the "right" control bits are provided. We refer to this bit pattern as *activation package*. On the other hand, the assignment of the activation package bits to the register only happens after the fabricated SoC is returned to the OEM from the foundry; when the SoC is in an untrusted facility (*e.g.*, foundry, assembly, or test), the activation package is not available. Furthermore, other bit patterns in addition to the activation package also correspond to other perfectly viable NoC topologies; indeed, without access to the activation package, there is no way to decide which one of the viable topologies was the one actually intended by the designer. We make this statement formal in Section 3.1. Correspondingly, a facility attempting to reverse-engineer the SoC does not have a way to determine the correct topology (among the viable ones) without access to the activation package.

## MUX Insertion Methodology

Our motivating example in Section 3.1 identifies the key ingredients of the ObNoCs methodology. ObNoCs takes an SoC design (typically in RTL) together with user directives for routers targeted for obfuscation. The output of the ObNoCs transformation is a transformed design with the routers obfuscated through redaction using the *MUX-DEMUX* switches as discussed above.

Algorithm 1 defines the procedure for inserting the MUX-DEMUX switch at each router $R$. Roughly, for each $S_i$ such that there is a link $R \rightarrow S_i$ and each $D_i$ such that there is a link $D_i \rightarrow R$, we instantiate a programmable MUX for each end of the router links. Here the function *RandomizeConnections* introduces non-determinism in connecting the outputs of a MUX, *e.g.*, in our motivating example, it will enable us to non-deterministically map the four candidate blocks to the four outputs of the DEMUX switch for $R_1$. Finally, for each MUX insertion, Algorithm 1 additionally records the bit pattern that must be loaded to the control register to recover the original topology. For the configuration of the MUX on the Source side as mentioned in Algorithm 1, each input of the generated MUX ($MUX_{R\_S_i}$) would be connected to all the signals of the Router outputs $R[S_i]$. The MUX output would be appended to a custom array MUX_out_list, which would serve as inputs to the Destination MUX's ($MUX_{R\_D_i}$). For the correct transformation involving the communication links, the MUX select lines have to be configured accordingly, for which the correct configuration for each Source MUX is stored in a variable $P_{Ri}$, with the input Source signal, the intended destination signal along with the connections of the source MUX ($in_i, S_i, D_i, MUX_{R\_S_i}[out]$), respectively. Using a *RandomizeConnections* function, the input signals of the source MUX are randomized, to increase the uncertainty of retracing in case of reverse engineering in a brute force attack scenario. The correct MUX configuration, using the select lines would be retrieved by the array index *i.e.* the position of the actual input signal by referring to the signals stored in $P_{Ri}$. The right select line configuration, $Sel_R$, is then appended to the comprehensive activation package ($Act\_Pkg[R]$) used to configure our entire transformation. The bit pattern required for the DEMUX switches (resp., MUX switches) at the output (resp. *input*) ports of router $R$ will be referred to as the *output* (resp. *input*) *activation package* for $R$. The bit pattern obtained by concatenating the activation packages of all routers

---

**Algorithm 1** Algorithm for *Custom MUX* Insertion

---

**Input**: Router $R$ with source and destination connections $R[S_i]$, $R[D_i]$

**Output**: Activation package for $R : R[P_i]$, *Act_Pkg*

***Source (DEMUX) Configuration***

  0: **for** single *Router connection* $R : R[S_i]$ , $R[D_i]$. $D_i \leftarrow S_i$

  0:    **for** all $S_j$ in $R[S]$, **do**

  0:      **Generate** $MUX_{R\_S_i}$.

  0:      MUX input :$MUX_{R\_S_i}[in_i] \leftarrow R[S_i]$,

  0:      MUX output :$MUX_{R\_S_i}[out]$

  0:      MUX_out_list **append** $MUX_{R\_S_i}[out]$

  0:      Store : $P_{Ri} = [in_i, S_i, D_i, MUX_{R\_S_i}[out]\ ]$

  0:      ***RandomizeConnections***$(MUX_{R\_S_i}[in_i])$

  0:      $Sel_R \leftarrow$ ***getIndex***$(MUX_{R\_S_i},\ in_i,\ S_i)$

  0:      *Act_Pkg*$[R]$ **append** $Sel_R$

  0:    **end for**

  0:    **for** all $D_j$ in $R[D_i]$, **do**

  0:      **Generate** $MUX_{R\_D_i}$:

  0:      $MUX_{R\_D_i}[in_i]$ **append** $P_{Ri}[MUX_{R\_S_i}[out]]$

  0:      $MUX_{R\_D_i}[out] \leftarrow R[D_i]$

  0:    **end for**

  0:  **end for**

  0: **return** *Act_Pkg*, MUX_out_list

     =0

***Destination (MUX) Configuration***

**Input**: Router $R$ with source and destination connections $R[S_i]$, $R[D_i]$, $MUX_{R\_S_i}[out]$, MUX_out_list.

**Output**: Activation package for $R : R[P_i]$, *Act_Pkg*

  0: **for** all *Router connections* $R_\backslash :$ , $R_\backslash[D_n]$.

  0:    MUX input :$MUX_{R_n\_D_i}[in_i] \leftarrow MUX_{R_n\_S_i}[out]$

  0:    $MUX_{R_n\_D_i}[in_i - 1] \leftarrow$ randomSelect(MUX_out_list, *size*-1)

  0:    MUX output :$MUX_{R_n\_D_i}[out]$

  0:    ***RandomizeConnections***$(MUX_{R_n\_D_i}[in_i])$

  0:    $Sel_{R_n} \leftarrow$ ***getIndex***$(MUX_{R_n},\ in_i\ ,\ MUX_{R_n\_D_i}[out]\ ,\ D_i\ )$

  0:    *Act_Pkg*$[R_n\ S_i]$ **append** $Sel_{R_n}$

  0:  **end for**

  0: **return** *Act_Pkg*

     =0

---

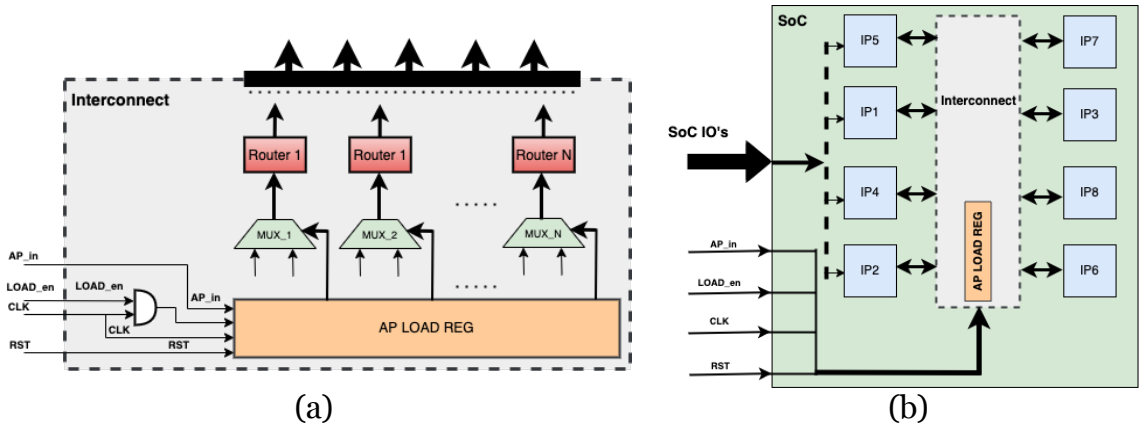

Figure 6: (a) Architecture of the Activation Package Loader Circuit. (b) Transformed SoC with Integrated Activation Package Loader.

in the NoC will be referred to as the *activation package of the NoC*, or simply *activation package* (*Act Pkg*).

**Remark 1.** *Algorithm 1 inserts MUXes on a port of router R such that the control configurations induce topologies involving blocks that are already connected to some port of R. However, note that the only requirement for functional correctness is that the intended topology is realized under the MUX control configurations that correspond to the correct activation package; no assumption is required for other configurations.* ObNoCs *can exploit this observation to connect R with other IPs creating additional topologies under different control configuration. This can provide additional protection against SAT attacks and brute force adversaries.*

## Activation Package Loader

An NoC design transformed by ObNoCs would realize the original topology when the registers connected to the controls of MUX and DEMUX switches are configured with the activation package. On the other hand, it is not possible to load the activation package directly on the fabricated SoC via parallel load from external inputs given the limited number of input pins available. ObNoCs correspondingly also introduces circuitry to load the activation package through a bit-shifting paradigm.

Fig. 6 shows the architecture for loading the activation package. The design is inspired by scan chain designs in VLSI testing. To streamline insertion, we create a single register bank, referred to as *activation package load register* (AP_LOAD_REG) for holding the activation package of the NoC. *AP_LOAD_REG* works on the Serial Input Parallel Output (SIPO) mechanism, where the activation package is provided one bit at a time per each clock cycle through the 1-bit primary input *AP_in*. The signal *LOAD_en* is gated with the clock and the bits are loaded into the shift register only when *LOAD_en* is asserted. Once the entire activation package is loaded into the register, the *LOAD_en* is then de-asserted and the parallel outputs of the register are connected to the respective programmable MUX select lines, with the correct configuration realizing the intended topology.

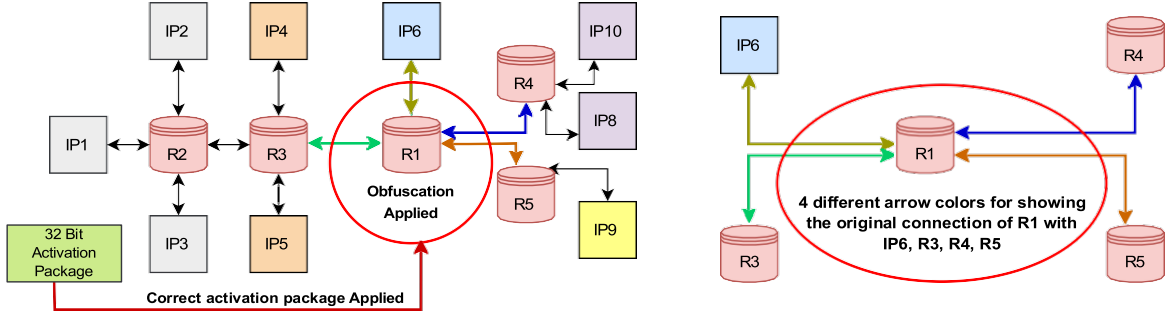

(a) The original topology.

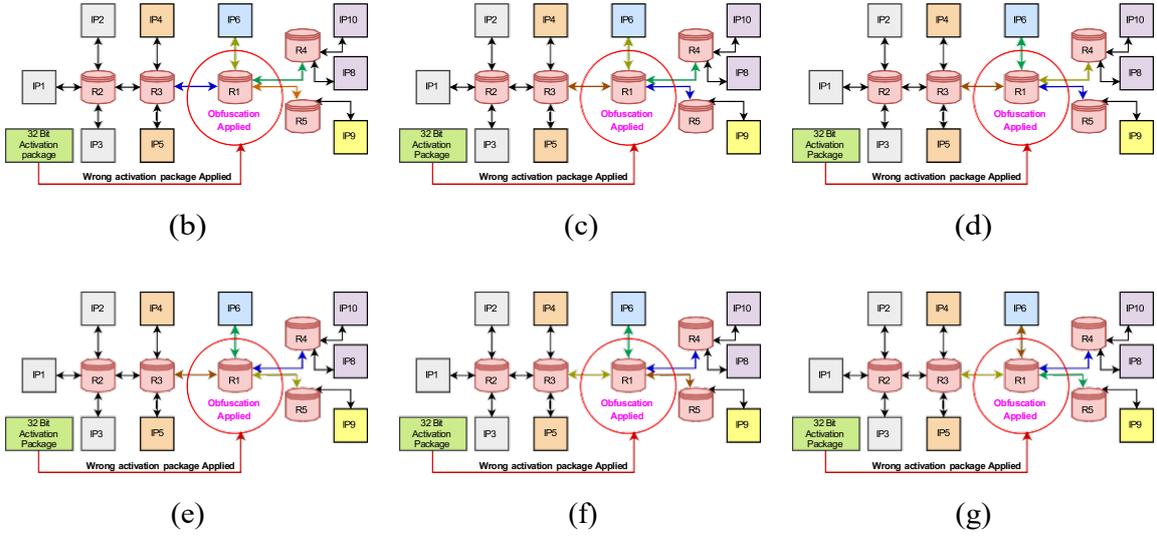

(b)  (c)  (d)

(e)  (f)  (g)

Figure 7: Different Functional Topology generation using ObNoCs. 7a indicates the original topology, while the other six topologies are legally functional, but none of them have the Original Functionality.

## Security Analysis

### Theoretical Guarantee

The key security guarantee of ObNoCs is that the original topology of the NoC interconnect cannot be derived from the obfuscated design without knowledge of the activation package. To make this statement formal, let us fix a design $D$ and let $D_o$ be an obfuscated design generated from ObNoCs with an $n$-bit activation package $A \in \{0, 1\}^n$. We call the topology graph $T$ of $D$ the ***intended topology***. Obviously, $N$ consists of the IPs and routers in $D$. We call any topology graph $G$ over $N$ a ***legal topology*** if for any node $v \in N$, the degree of $v$ in $T$ is the same as the degree of $v$ in $G$. Given any binary string $b \in \{0, 1\}^n$, we refer to graph $G_b$ derived from $D_o$ by setting the MUX controls to be the bitstring $b$ as the *topology of $D_o$ induced by $b$*. Obviously, the topology of $D_o$ induced by the activation package $A$ is the intended topology. Then the following property defines the security guarantee of ObNoCs.

**Legality Enumeration.** Given a binary string $b \in \{0, 1\}^n$, let $m$ be the average degree in the topology of $D_o$ induced by $b$. Then the total number of unique legal topologies is $m!$.

The property is a straightforward from the ObNoCs construction and the observation that the connections induced by the MUX and DEMUX switches at the input and output of each router respect the original in-degree and out-degree of each node. A consequence of the property is that the intended topology $T$ is different from all the other legal topologies induced by the obfuscated design $D_o$ only by the specifics of the bitstream that defines the activation package $A$. It follows that an adversary without access to $A$, cannot distinguish the intended topology from other legal topologies.

**Remark 2.** *Informally, the* intended topology *here is really the topology that the designer desires, while all topologies that can be realized through some values of the configuration bitstream are legal topologies. It follows that the intended topology is simply one of the legal topologies, in particular the specific legal topology that corresponds to the activation package that the designer has in mind. However, other than the specifics of the activation package there is nothing else to distinguish the intended topology from other (unintended) ones. In particular, each legal topology would constitute an SoC in which IPs could communicate with one another. Note that some values of the configuration bitstream would result in a* non-functional *topology in the sense that the same port of two IPs is mapped to the same output. Fig. 7 shows some possible unintended and non-functional topologies, together with the (unique) intended topology. It is obvious that an adversary can detect non-functional bit configurations; however, a large number of legal topologies and the uniqueness of the intended topology among the legal ones implies that identifying the intended topology is equivalent to identifying the activation package.*

To explain the consequences of the above analysis a bit further, consider the SoC design shown in Fig. 4 For each router obfuscation, we used 16 4x1 MUXs. Since each 4x1 MUX requires two select lines, the total number of bits for the activation package is 32. However, not every bit pattern corresponds to a unique legal topology, due to the recombination of the combinational logic involved. Even if only one router is obfuscated for the design in Fig. 4, the number of legal topologies is 4! = 24. Fig. 7 illustrates 7 of the 24 possible topologies (including the intended one) whereas Fig. 7a indicates the intended topology upon insertion of the correct activation package. The four different colored arrows in the red encircled portion with router $R_1$ indicates the original connection with the $R_3$, $R_4$, $R_5$ and $IP_6$. These connections from $R_1$ can be controlled by our ObNoCs which has been depicted in the rest of the figures in Fig. 7. On the other hand, there is no *a priori* reason to determine which one of the legal topologies is the intended one (from a functional standpoint) other than the "idiosyncrasies" of the designer's choice. Furthermore, as our empirical evaluation shows (see below) the different legal topologies induce very different computations; consequently, the adversary cannot get away with identifying a topology close enough to the intended. Finally, the number of possible legal topologies can be compounded by simply composing the obfuscation in stages: for a 2-stage MUX-based obfuscation from Fig. 5, there are 4! · 4! = 576 legal topologies.

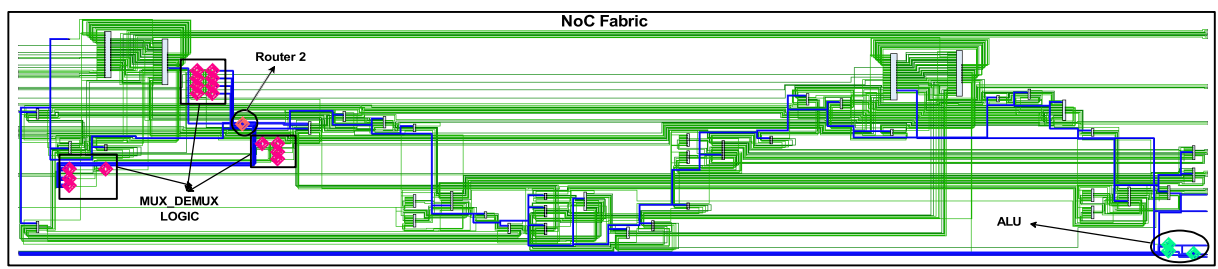

Figure 8: Schematic of an NoC interconnect in Xilinx Vivado along with an ALU

We conclude this section with a brief comment on the storage of the activation package. Recall that our threat model considers an untrusted foundry or testing facility that has access to the layout information for the unfabricated SoC as well as silicon implementation of the fabricated SoC. Since the topology information is redacted by ObNoCs through the MUX insertion and the control bits of the MUX are only programmed subsequently with the activation package by the OEM, neither the layout nor the fabricated SoC has the activation package stored in the chip as long as the SoC is in control of the untrusted entity. However, when the SoC goes to field, it includes the activation package (which is obviously required to make the SoC functional). In this paper, we do not specifically discuss the storage of the activation package in field since it is outside the scope of our threat model. However, ObNoCs does not need any specialized mechanism for storage and application of activation package; any mechanism for storing keys used for unlocking an obfuscated circuit (*e.g.*, through logic locking) can be applied for this purpose, *e.g.*, a typical mechanism involves utilizing a small tamper-proof ROM in the design to hold the key bits.[1]

## Resiliency Against Known Attacks on MUX-based Obfuscation

MUX-based transformation is a common strategy for obfuscating hardware functionality. Correspondingly, there has been significant research on attacks against such obfuscation. To demonstrate the effectiveness of ObNoCs we consider some common attack strategies

---

[1]Current state-of-the-art locking techniques generally use an on-chip ROM although techniques for storing the key in the cloud together with other provisioned assets is being explored. Generally, the requirement is that the storage should be tamper-proof and communication of the key for its application should satisfy non-observability requirements.

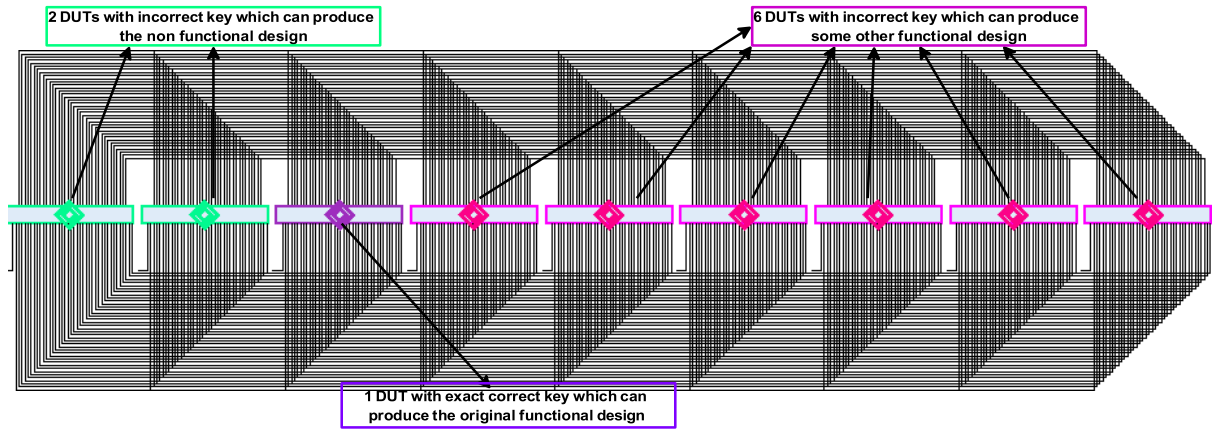

Figure 9: Multiple DUTs for the same Obfuscated NoC to compare the output for different activation package combination

and explain how ObNoCs provides resiliency against these attacks.

**Remark 3.** *Note that there has been no previous work to our knowledge on obfuscation of NoC topologies. Although traditional logic locking techniques can be applied to SoC designs with NoC fabrics, they are generally applicable only to gate-level netlists. Performing them on a complete SoC, while possible in principle, incurs significant computational overhead and is infeasible in practice. Indeed,* ObNoCs *avoids netlist designs works on RTL-level designs to avoid such computational overhead in obfuscation. Furthermore, the presence of encrypted IPs in Quartus makes it practically infeasible to extract a complete netlist from Quartus to enable a direct comparison with other obfuscation techniques. Furthermore, given these factors, a direct implementation and evaluation of traditional attacks (which also apply to gate-level netlists) could not be performed as part of direct resiliency evaluation. However,* ObNoCs *architecture in spite of being a MUX-based framework, is robust against typical attacks on MUX-based obfuscation techniques as explained below.*

**ML-based Attacks** Machine Learning attacks on obfuscation [4,5] use a variety of ML techniques to predict MUX configurations, thereby inferring unbofuscated links, *e.g.*, Al-rahis *et al.* [5] use GNN to perform such predictions. However, the *success* of the attacks relies on the fact that the obfuscation is performed on a netlist *post synthesis*. Correspondingly, the ML strategies are employed on unobfuscated links indicating dependence on the technology library. However, in ObNoCs, this assumption is broken since the obfuscation is performed *before synthesis* and MUXs used are architectural entities with no dependence on the underlying technology library for insertion.

**SAT Attacks** SAT attacks [18] form another popular class of attacks to break hardware obfuscation. They depend on an oracle (*e.g.*, a fabricated design on which one can execute test patterns) to answer the sequence of SAT queries. The SAT queries are computed to enable reconstruction of specific bits of the activation package. Resiliency against SAT attacks depends on the *query complexity*, *i.e.*, the number of SAT queries involved in the attack. For ObNoCs, the query complexity is defined by the number of legal topologies derived in Section 3.1. Furthermore, recall that ObNoCs enables extensibility by extending the number of router connections with additional IPs, which can be applied to achieve additional resiliency to SAT attacks. Finally, since SAT attack (and other attacks) cannot distinguish the intended topology from other legal topologies, the correct activation package even if produced as a candidate bit pattern through SAT attack, cannot be vetted as indeed inducing the intended topology (see below).

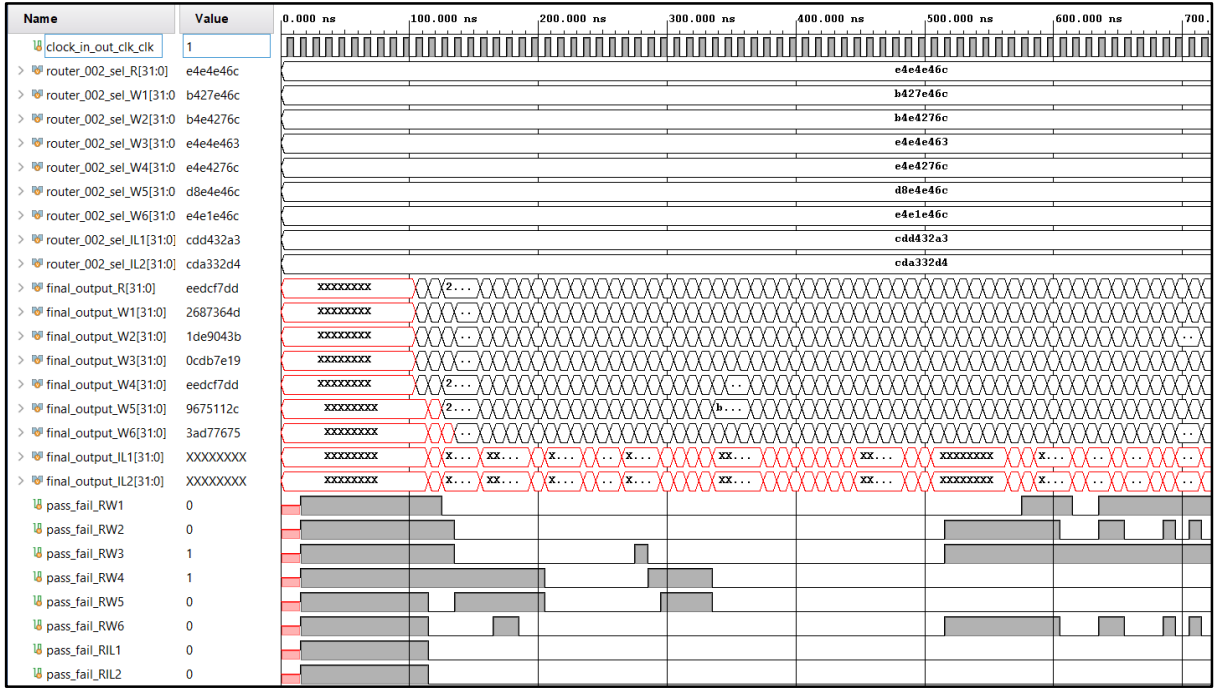

Figure 10: Comparison of the Output of all 9 DUTs on nine different activation package insertion

**Specific Attack Instances** We have analyzed the resilience of ObNoCs on two specific attack instances: (1) redundancy-based attack strategy [14], and (2) the SNAPSHOT deep learning attack [17]. The key idea of redundancy-based strategy is that an incorrect activation package would lead to incorrect functionality, which is exploited through different redundancy levels to infer the correct key. SNAPSHOT relies on a deep learning strategy to predict the correct activation package that corresponds to "intended" functionality. However, for ObNoCs, legal topologies that differ from the intended one would still produce a functional SoC, simply not the one intended by the SoC designer. Consequently, since the technique relies on functional validation to detect incorrect behavior, it cannot distinguish counterfeit SoCs with legal topologies from the intended one.

**Remark 4.** *Obviously, any of the strategies above can find legal topologies. This leaves*

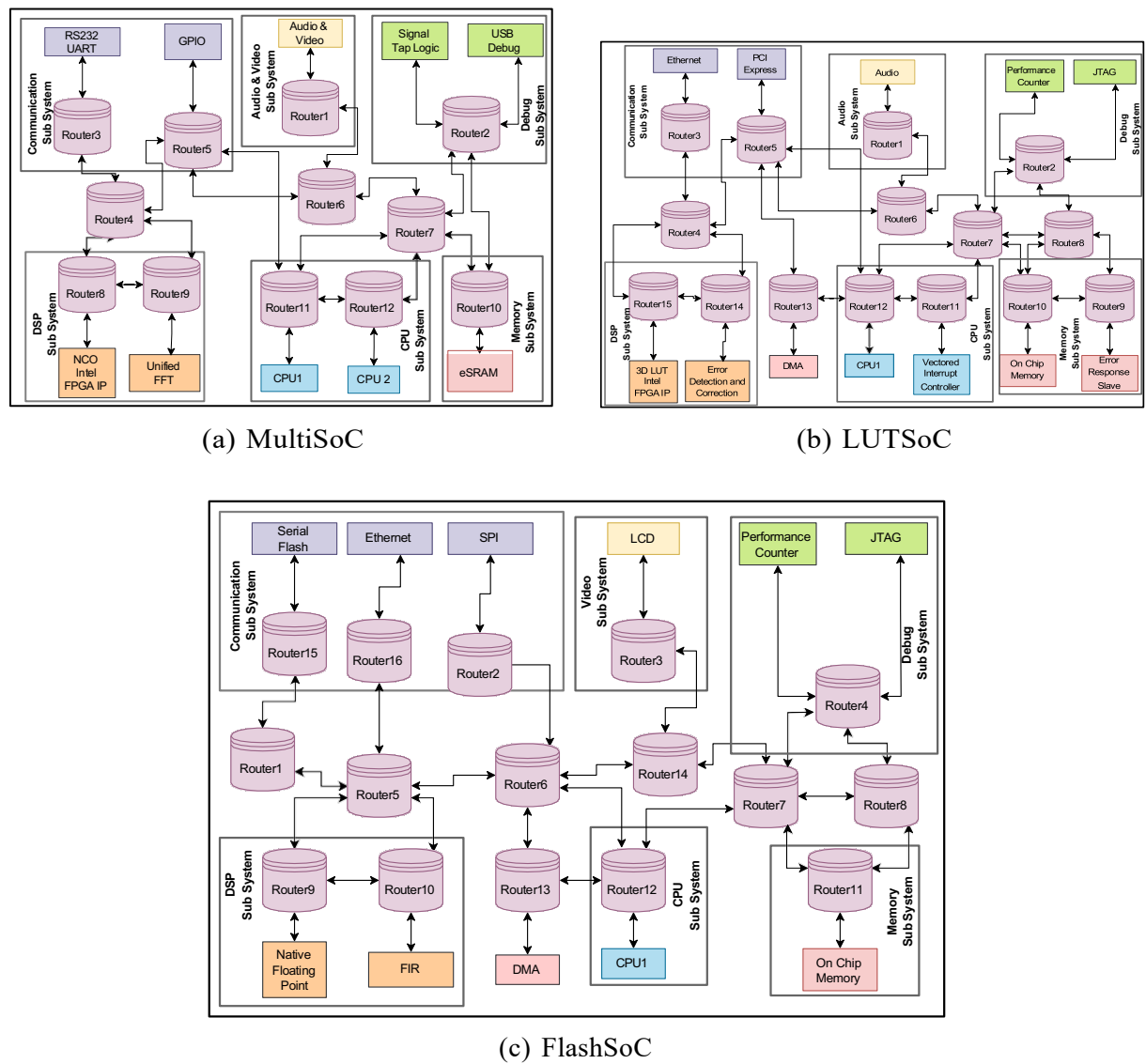

(a) MultiSoC      (b) LUTSoC

(c) FlashSoC

Figure 11: Three SoC designs used for ObNoCs evaluation. All three designs have been implemented on Quartus Prime with Platform Designer.

*open the possibility that the adversary can simply enumerate (and fabricate) all counterfeit SoC variants corresponding to legal topologies. This is indeed a threat if the number of legal topologies is small. However, as discussed above,* ObNoCs *can apply staging strategy to expand the space of legal topologies by multiplicative factors. Note that adding a single stage would multiply the number of legal topologies by a factor of* 24*. Indeed, simply using the 2-stage strategy of Fig. 5 with* 576 *possible legal topologies makes it infeasible for the adversary to fabricate all the SoC variants.*

## Empirical Analysis

The security analysis above shows that it is infeasible for the adversary to reverse-engineer the intended topology from ObNoCs obfuscation. One objection to that thesis could be that many topologies could be functionally similar, so a counterfeit SoC that does not create the intended topology could still serve as a valid counterfeit. To evaluate this possibility, we considered the a NoC interconnect and created a simulation environment with an ALU module integrated for effective analysis of the functionality. Fig. 8 shows a Xilinx Vivado setup for this empirical analysis.[2] In this figure, the router including all the required MUXes for the obfuscation and the ALU module have been pointed out for a better understanding of the design. The key idea is to drive the interconnect with different workloads corresponding to inter-IP communications and observe the functional impact on the computation of the ALU. The simulation used a total of 9 DUTs (Fig. 9) of which only one is driven by the correct activation package while the rest are incorrect. Fig. 10 shows the result of the simulation. The correct functionality is derived only with the correct activation package. In the simulation we have provided 9 different activation package for 9 DUTs. On the other hand, the ALU is still functional (but incorrect and different from the functionality in the intended topology) in 6 DUTs, which correspond to legal topologies different from the intended one. Note that functional validation of the ALU would consider all 7 topologies to be correct, thwarting attacks that depend on functional validation.

Recall that the theoretical analysis above showed that the obfuscated design can correspond to a number of different legal topologies of which only one is the intended one. To substantiate this, we have created a simulation environment in Xilinx Vivado. For this simulation, we have taken only the interconnect generated in Platform Designer as our Design Under Test (DUT).

The interconnect has been shown in Fig. 8 and we have integrated one ALU module in it for better analysis. The NoC interconnect fabric is responsible for all the data communication for the SoC. To check data transfer from the NoC, we have taken the ALU result as our reference. In our simulation, there are total 9 DUTs in Fig. 9 where only one of them is driven by the correct activation package, 8 are driven by the incorrect activation package.

## Overhead Analysis

### Benchmark SoCs

A key challenge in research on SoC security validation is the lack of an appropriate testbed for evaluation. Evaluation of ObNoCs requires SoC designs with a realistic NoC-based communication fabrics connecting disparate IP cores that reflect realistic design complexity. Unfortunately, there is no open-source SoC design satisfying these requirements.

---

[2]Our primary design and experimental testbed is Quartus, which we used for overhead analysis as discussed in Section 3.1. In this figure, the router including all the MUXes used for this obfuscation, and the ALU module have been highlighted for a better understanding of the design. However, for the empirical security analysis, we used Vivado for ease of simulation of different instances. The interconnect model was exported from Platform Designer to Vivado to perform this simulation.

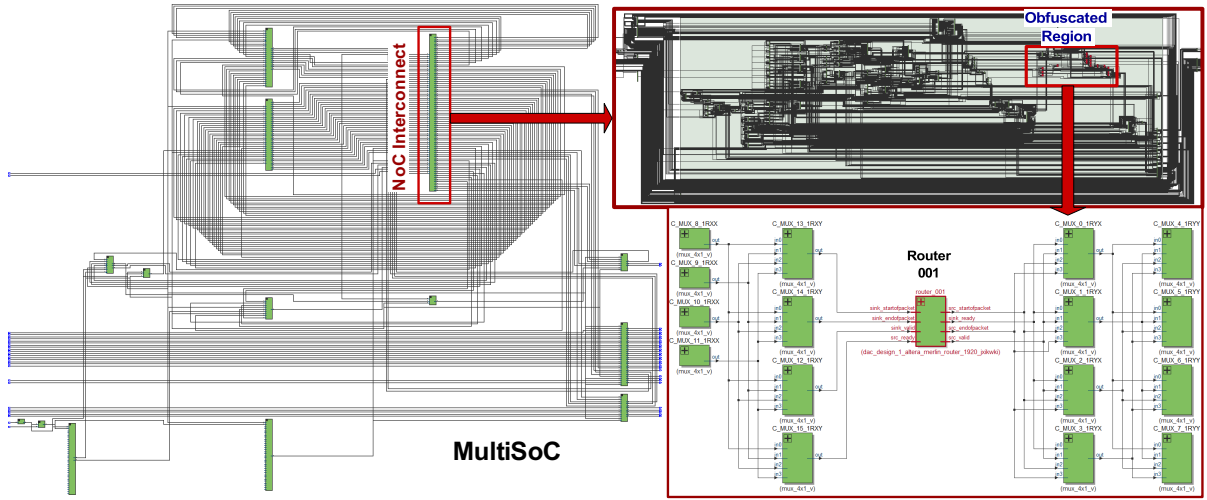

Figure 12: Schematic of Benchmark SoC in Fig. 11(a) from Quartus along with the obfuscated Interconnect using ObNoCs technique

To address this problem, we have developed our own SoC benchmarks, shown in Fig. 11. We use three SoCs with different interconnect sizes to ensure that our results are generalizable. The three SoCs are derivatives of each other, comprised of similar IPs but with different interconnect topologies. We refer to the SoCs as *MultiSoC*, *LUTSoC*, and *FlashSoC* respectively. Each SoC is comprised of six subsystems, with each subsystem containing specific IPs; The CPU subsystem is made up of two NIOS processors; the memory subsystem consists of an on-chip memory and a DMA controller; the Communication subsystem includes an SPI module, an Ethernet, and a Serial Flash module; and the Debugging subsystem consists of a JTAG and a Performance Counter IP.

While undoubtedly simple compared to an industrial product, the SoCs are non-trivial; the RTL for the unobfuscated (original) SoC for each variant runs to about $100,000$ lines of RTL, with the interconnect fabric accounting for $10,500$ lines. The Block Memory bits for the SoCs are $3.6 \times 10^5$, $9.8 \times 10^5$, and $1.2 \times 10^6$ respectively, and there are respectively 12, 15 and 16 routers in the NoC fabrics.

## Platform and Implementation Details

We implemented ObNoCs using the Intel Quartus Pro 21.3. For our SoC design and integration, we have used the Platform Designer frontend of Quartus and implemented the ObNoCs methodology for systematically redacting various router connections as reported below. For evaluating ObNoCs, we have designed the SoCs using the Platform Designer (PD) frontend of Quartus Pro 21.3 and implemented on the Agilex F Series FPGA, AGFA012R24A2E2V. Fig. 12 shows the schematic diagram of the obfuscated SoC where the enclosed red rectangle indicates the NoC interconnect of the SoC after transformation by ObNoCs.

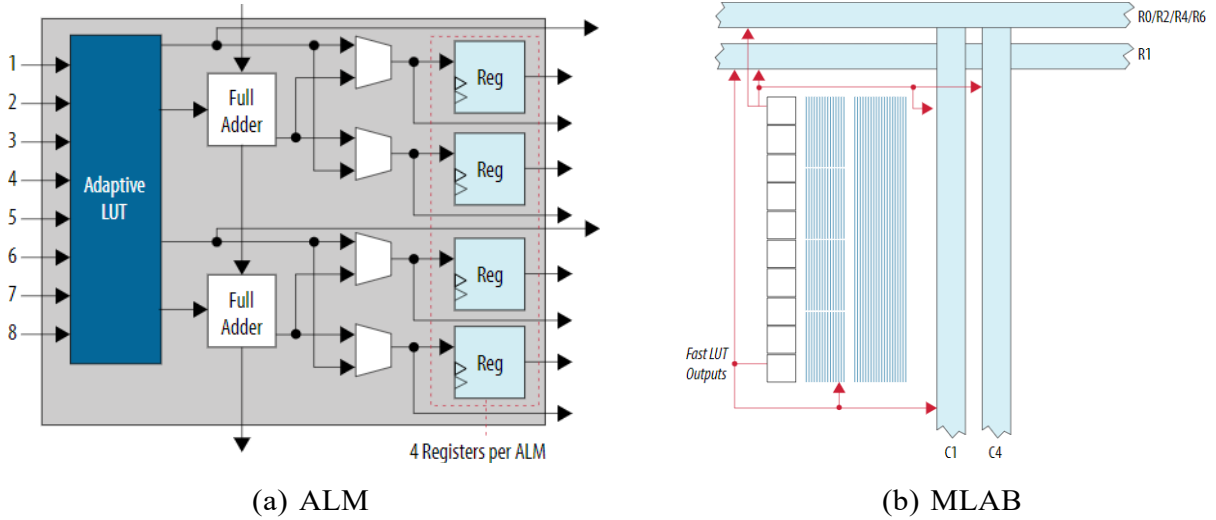

(a) ALM                              (b) MLAB

Figure 13: ALM and MLAB Architecture in Intel® Agilex™ 7 FPGAs

Table 1: Obfuscation Level based on No. of Routers

| Obfuscation Level | 0 | I | II | III | IV |
|---|---|---|---|---|---|
| No. of Routers | 0 | 2 | 4 | 8 | 16 |

## Result Analysis

To provide a fine-grained analysis of the overhead induced by our interconnect transformation, we define five levels of obfuscation (0-IV) as shown in Table 1 where obfuscation level 0 indicates no obfuscation, *i.e.*, the original design. Estimates of power and hardware resource overheads have been derived from the Quartus platform. Quartus provides three different metrics on the resource overhead: (1) power consumption; (2) number of required Adaptive Logic Modules (ALMs) and (3) number of Logic registers (which estimates the state elements). The Intel Agilex device consists of Logic Array Blocks (LABs) shown in Fig. 13b and is implemented by enhanced ALM modules mentioned in Fig. 13a which allows efficient implementation of logic functions. Each ALM consists of two Adaptive Look-up Tables (ALUTs), two dedicated embedded Adders and four dedicated Registers [3]. We have followed the user guidelines from [2] to carefully analyze the resources from Quartus generated reports.

Figs. 14a, 14b and 14c show the resource utilization for different levels of obfuscation. Since for each obfuscation level there are variations in utilization depending on which specific routers are selected as obfuscation targets (see below), we consider the average utilization for each combination, *e.g.*, the resource overhead for Obfuscation Level I is calculated by considering the $\binom{n}{2}$ results for designs derived by obfuscating each possible

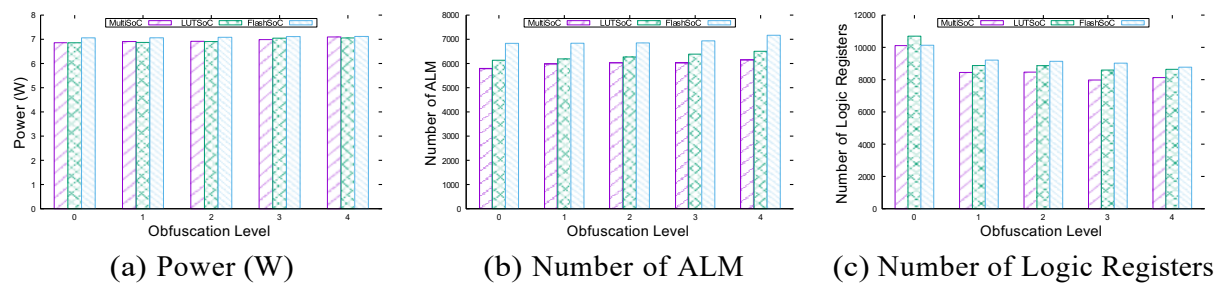

(a) Power (W)    (b) Number of ALM    (c) Number of Logic Registers

Figure 14: Resource Analysis for *MultiSoC*, *LUTSoC* and *FlashSoC* at different obfuscation level

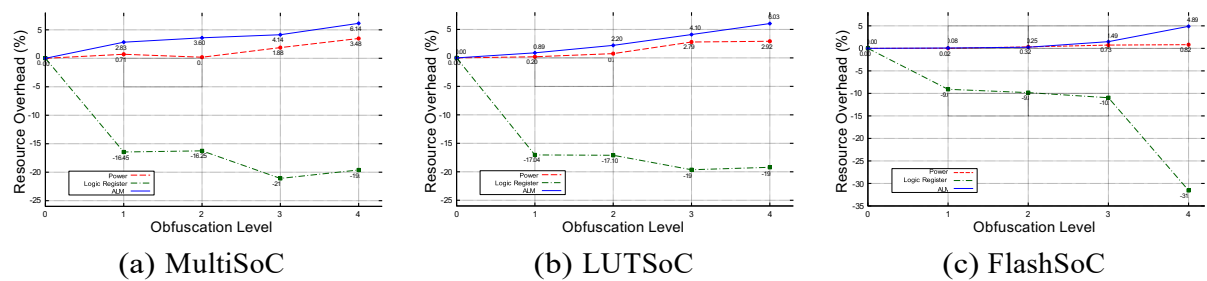

(a) MultiSoC    (b) LUTSoC    (c) FlashSoC

Figure 15: Resource Overhead Rate for all 3 SoCs at different obfuscation levels.

pair of routers and considering the arithmetic mean where $n$ is the number of routers available inside the NoC fabric. Power and number of ALM both are increasing with the obfuscation level for all three SoCs *MultiSoC*, *LUTSoC* and *FlashSoC*. Note that the power and number of ALM units are increasing with the obfuscation level increment whereas the number of logic registers are decreasing.[3]

Figs. 15a, 15b and 15c show the resource overhead rate results corresponding to various obfuscation levels for the three designs. Note that this results in an average of only 5.68% overhead increase in the number of ALM modules, 2.41% overhead increase in power consumption, 23.46% reduction in usage of logic registers, where all of the three resource parameters have been compared with their respective unobfuscated designs. Nevertheless, we can conclude that the addition of logic by ObNoCs appreciably impacts very little in resource overhead.

It is interesting to understand the decrease in the logic register overhead which can be attributed to the increase in the number of ALMs, with each ALM consisting of 4 additional registers. Since the obfuscation technique is implemented using pure combina-tional logic, the primary effect of the overhead is an increase in the number of ALMs as well as total power. This combinational logic is implemented with ALMs, which leaves fewer resources available to build logic registers. In general, all of the design's logic functions, including both combinational and sequential logic, must share the resources in an FPGA, such as ALMs and flip-flops. Additionally, the number of flip-flops that can be used to build the registers are constrained if the combinational logic calls for a large number of inputs or outputs. Furthermore, additional combinational logic introduced during obfuscation (*e.g.*, MUX and DEMUX structures) permits Quartus to perform aggressive optimizations in logic synthesis.

---

[3]We have taken the ALM and Logic Registers values obtained from Quartus synthesis report to represent area overhead. Note that it is not possible to export the netlist synthesized from Quartus to an external environment (*e.g.*, Xilinx Vivado or Synopsys Design Compiler) since the Quartus environment integrates a number of encrypted IPs. Consequently, area estimation using those traditional tools is not possible. For an ASIC implementation, the overhead result can be different.

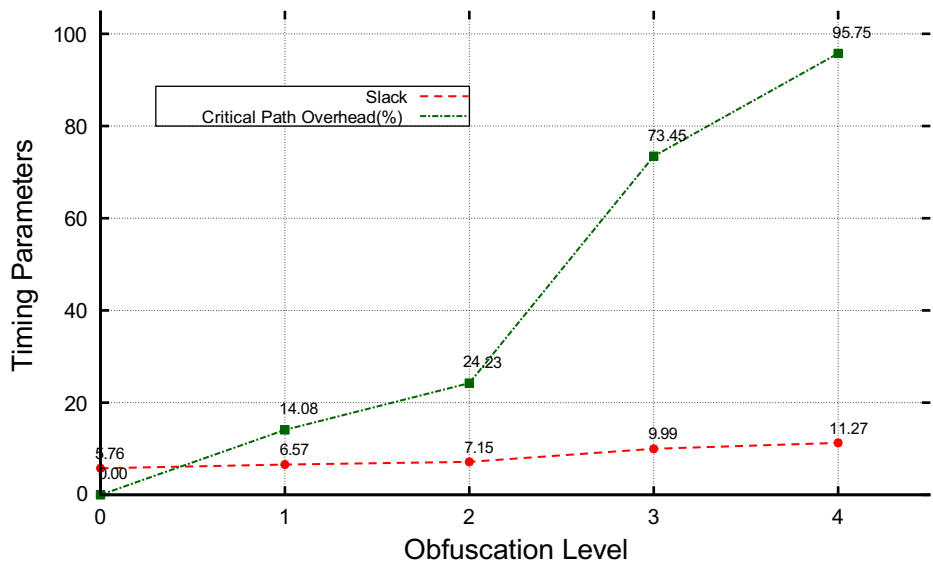

Figure 16: Timing Analysis for *LUTSoC* at different obfuscation levels.

Finally, the additional hardware logic introduced for obfuscation may affect the critical path of the system. From Fig. 16, we observe that the critical path increases due to the obfuscation level. This suggests that the increased security due to the increasing degree of obfuscation induces a trade-off by adding combinational gate delay. However, the delay induced is significant only for obfuscation levels III and IV. Furthermore, in SoCs with a larger number of IPs, the combinational delay in the interconnect would be proportionately less. This timing resource analysis is solely based on Quartus Timing Analyzer report [2] with considering slack and critical path as timing parameters.[4].

---

[4]Our observations are based on pre-synthesis estimates, which can be different from post-synthesis implementation. Implementing ObNoCs in FPGA and considering the correlation between pre-synthesis and post-synthesis timing correlation will be considered in future work. However, such correlations are not germane to ObNoCs and follow standard practice of timing correlations between various timing models used pre-synthesis vis-a-vis timing on actual artifacts

## 3.2 Challenges in NoC Obfuscation: Pre-Synthesis vs. Post-Synthesis

### Limitations of Pre-Synthesis Obfuscation

Pre-synthesis obfuscation techniques operate on the RTL description. However, this technique has several limitations that can impact its effectiveness and practicality. The primary limitation is that the synthesis modifies the structure and connectivity, making the obfuscation technique unusable. In Fig. 17 we show a design and their corresponding obfuscated design for both pre-synthesis and post-synthesis. All the router IOs with associated IPs are considered for applying the obfuscation in pre-synthesis. In the post-synthesis obfuscated design, we see many signals are missing compared to the pre-synthesis obfuscated design. Fig. 18(a) and (b) show a practical router connection with the associated IPs inside an NoC fabric for both pre-synthesis and post-synthesis. There are various signals among the *router_000* and connected IPs such as *sink_data*, *sink_endofpacket*, *sink_startofpacket*, *sink_valid*, and *src_ready*. This post-synthesis alters the structure and connectivity of this design. The synthesized gate-level netlist only has the *sink_data* signal and *src_valid* signal available to the *router_000*. It also changes the data width for various signals. For example, the *sink_data* signal has a width of 130 bits in the pre-synthesis, whereas this signal only contains 11 bits in the post-synthesis. The logic optimization also affects the design signals introduced by the obfuscation algorithm. Indeed, we implemented the obfuscation algorithm of the experimental analysis of Halder *et al.* on the same SoC that we use to validate POTENT in Section 3.3. Note that the obfuscation architecture itself was shown by previous work to be provably secure against reverse-engineering attacks, including SAT attacks. *However, we could successfully apply the SAT attack on the synthesized gate-level netlist after logic optimization and find the correct key within 10 seconds.* On the other hand, the SAT attack does not succeed on the obfuscation performed by POTENT (on the same netlist) as illustrated in Section 3.3.

### Challenges in Post-Synthesis Obfuscation

Post-synthesis obfuscation, which operates directly on the gate-level netlist, can avoid the challenges associated with pre-synthesis but introduce a new set of challenges. One primary challenge is the increased complexity of the gate-level netlist compared to the RTL description, making it difficult to identify and isolate specific parts of the NoC for obfuscation without disrupting the design's functionality or performance [8]. Obfuscation techniques should preserve the original functionality and performance while minimizing overhead in terms of area, power, and latency. This requires careful analysis and optimization of the modifications introduced during the obfuscation process [16]. Another challenge is resilience against attack scenarios such as functional and structural analy-sis [19]. The obfuscation methodology must introduce sufficient ambiguity and complexity into the netlist to hinder attackers from inferring the original NoC topology, communication paths, and connected IP cores.

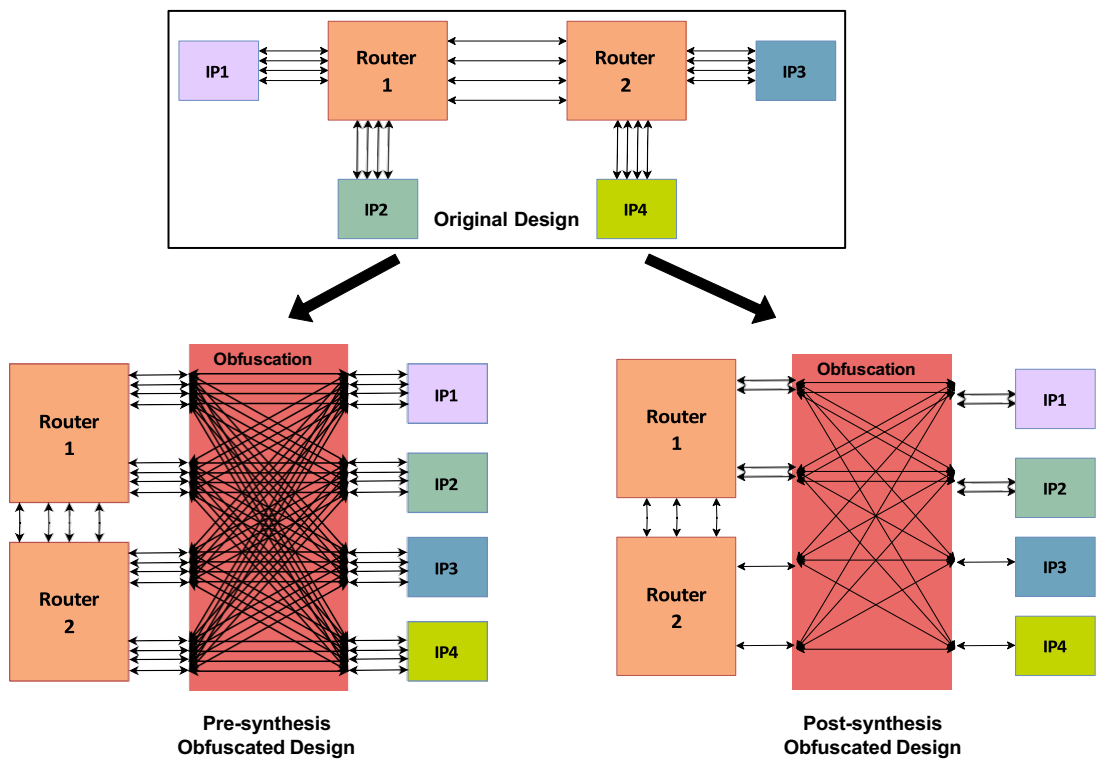

Figure 17: Pre-Synthesis vs Post-Synthesis Obfuscation Analysis

## 3.3   POTENT  Framework

The POTENT framework consists of two stages. In that first stage, a connectivity matrix is created, which serves as the main building block for POTENT. With the help of this matrix, we integrate the obfuscation switches into the router locations of the NoC. The second stage involves developing the obfuscation switch, which is integrated into the gate-level netlist according to the connectivity matrix.

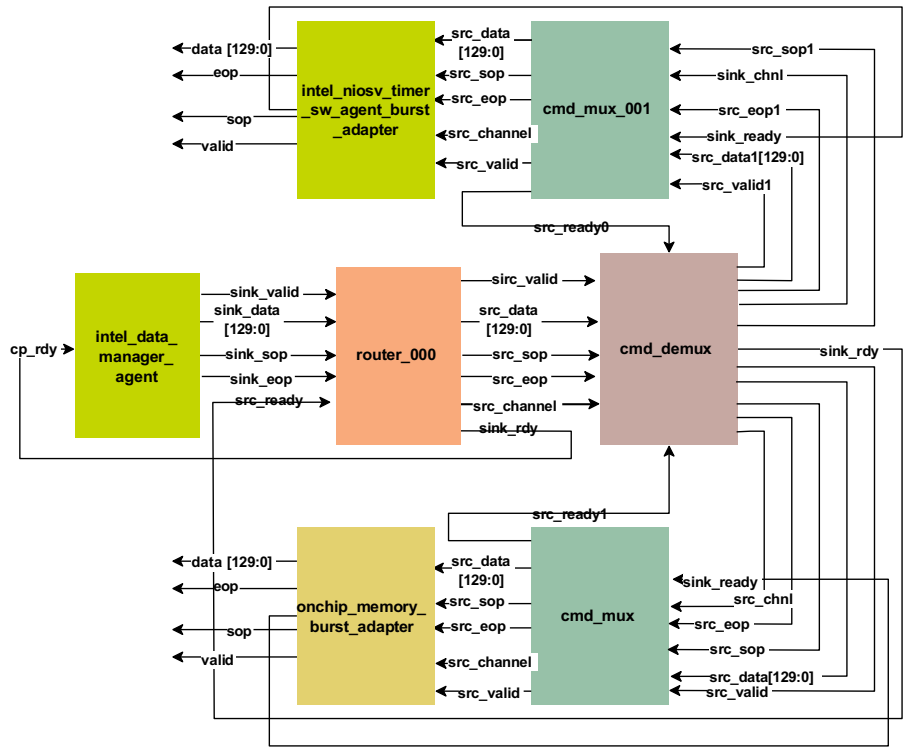

(a)

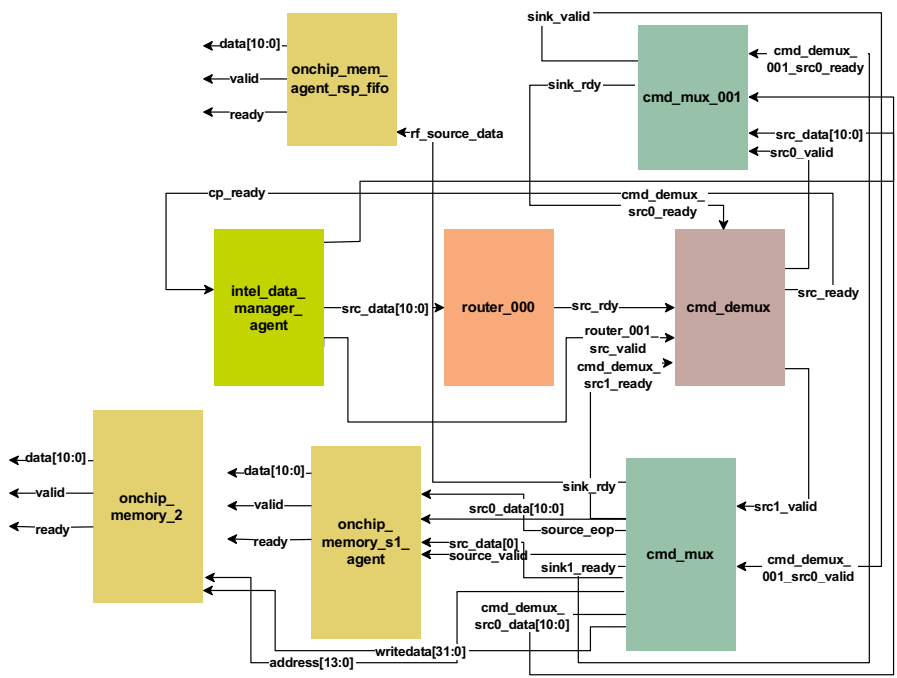

(b)

Figure 18: Analysis of a Single Router Connection inside a NoC fabric (a) Pre-Synthesis (b) Post-Synthesis

---

**Algorithm 2** Connectivity Matrix Generation

---

**Require:** Router $R$ with input $R_i$ and output $R_o$

**Ensure:** Connectivity Matrix $M_{\text{final}}$

1: Initialize $M_{\text{pre}}$ and $M_{\text{post}}$ as zero matrices for pre- and post-synthesis connections.
2: **for** $i = 1$ to $|R_i|$ **do**
3:     **for** $j = 1$ to $|R_o|$ **do**
4:         Set $M_{\text{pre},ij} = 1$ if input $i$ is connected to output $j$ in pre-synthesis, else $M_{\text{pre},ij} = 0$
5:         Set $M_{\text{post},ij} = 1$ if input $i$ is connected to output $j$ in post-synthesis, else $M_{\text{post},ij} = 0$
6:     **end for**
7: **end for**
8: **Connectivity Matrix Generation for Router $R$:**
9: Initialize matrix $M_{\text{final}}$ with dimensions $|R_i| \times |R_o|$
10: **for** $i = 1$ to $|R_i|$ **do**
11:     **for** $j = 1$ to $|R_o|$ **do**
12:         **if** $M_{\text{pre},ij} = 1$ AND $M_{\text{post},ij} = 1$ **then**
13:             Set $M_{\text{final},ij} = 1$ indicating the signal is preserved in both
14:         **else**
15:             Set $M_{\text{final},ij} = 0$
16:         **end if**
17:     **end for**
18: **end for**
19: **return** Connectivity Matrix $M_{\text{final}}$ for Router $R$ =0

---

## Connectivity Matrix Generation

In a SoC, an NoC fabric contains multiple routers which are connected to different IPs. In our benchmark design, there are a total of 6 routers inside the NoC fabric. For each router, a connectivity matrix is generated based on the pre-synthesis and post-synthesis netlist information. Fig. 18 shows a subset of the NoC fabric with *router_000* and the corresponding connections of this router with the IPs for both the pre-synthesis and post-synthesis netlist.

The connectivity matrix generation process, as outlined in Algorithm 2, takes the router $R$ with its input $R_i$ and output $R_o$ as input and generates the final connectivity matrix $M_{final}$. The algorithm first generates connectivity maps for the router $R$, initializing $M_{pre}$ and $M_{post}$ as zero matrices for pre- and post-synthesis connections, respectively. It then sets the values in these matrices based on the connections between the inputs and outputs in the pre-synthesis and post-synthesis netlist. Finally, the algorithm generates the connectivity matrix $M_{final}$ by comparing the pre-synthesis and post-synthesis

Table 2: Connectivity Matrix for **router_000**

| Connected IP1 (*intel data manager agent*) | | Targeted ROUTER (*router 000*) | | Connected IP2 (*cmd demux*) | |
|---|---|---|---|---|---|
| **Output** | **Connected Wire** | **Input** | **Output** | **Connected Wire** | **Input** |
| align address to size128 | N43845[14] | align address to size | reduce nor 010 | N43857 | router |
| align address to size129 | N43845[13] | align address to size0 | | N43851[2] | write addr data both valid |
| align address to size130 | N43845[12] | align address to size1 | | N43851[1] | sinko ready |
| align address to size131 | N43845[11] | align address to size2 | | N43851[0] | cmd mux 001 |
| align address to size132 | N43845[10] | align address to size3 | | | |
| align address to size133 | N43845[9] | align address to size4 | | | |
| align address to size134 | N43845[8] | align address to size5 | | | |
| align address to size135 | N43845[7] | align address to size6 | | | |
| align address to size136 | N43845[6] | align address to size7 | | | |
| align address to size137 | N43845[5] | align address to size8 | | | |
| align address to size138 | N43845[4] | align address to size9 | | | |

connectivity maps, which are shown in Table 2 for *router 000*.

## Obfuscation Switch Generation and Integration

The core of POTENT lies in the obfuscation switch, $\Omega$, which is integrated into each NoC router. For a router with $n$ ports, $\Omega$ employs a permutation-based obfuscation approach to reconfigure signal paths dynamically.

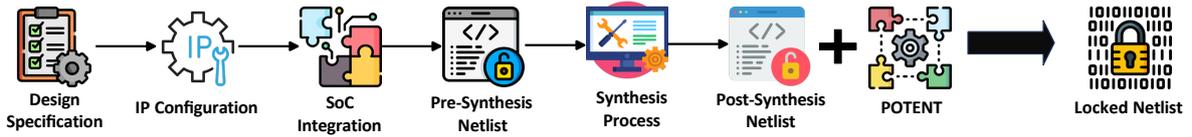

Figure 19: POTENT Integration Flow

The set of possible signal connections, $S$, undergoes permutations, generating $n!$ mappings. Each mapping is associated with a unique key from the key space $K$, a binary vector space of dimension $b$, resulting in $|K| = 2^b$ potential keys. The relationship between keys and signal mapping is $P(k_i)$.

$$P(k_i) = \sigma_i, \quad \forall i \in \{1, 2, \ldots, n!\}$$

Here $\sigma_i$ represents the $i$-th permutation of signals within $S$. As the number of routers $N_R$ increases with the number of IPs, the dimension of $K$ grows exponentially, providing a large space to defend against attackers. The total key space required for $N_R$ routers is given by $|K_{\text{total}}| = 2^{(N_R \cdot b)}$. The integration of $\Omega$ across the routers helps to create multiple unique network topologies under each unique key, resulting in an updated connectivity matrix, $M_{\text{final}}^{(R^{\text{obf}})}$.

$$M_{\text{final},ij}^{(R_{\text{obf}})} = \begin{cases} 1 & \text{if } P(k)(i, j) \text{ yields a valid} \\ & \text{connection under } \Omega, \\ 0 & \text{otherwise} \end{cases}$$

Algorithm 3 describes the process of generating and integrating $\Omega$ with the router $R$

to form the obfuscated router $R_{obf}$. The algorithm takes the connectivity matrix $M_{final}$ from Algorithm 2 as input and outputs an obfuscated router $R_{obf}$.

---

**Algorithm 3** Obfuscation Switch Generation and Integration

---

**Require:** Connectivity Matrix $M_{final}$ from Algorithm 2
**Ensure:** Obfuscated Router $R_{obf}$
  1: ***Generate Obfuscation Switch*** $\Omega$***:***
  2: Let $S$ be the set of all signals in $M_{final}$
  3: Define $P$ as the set of all permutations of signals, $P : S \rightarrow S$, where $|P| = 4! = 24$
  4: Assign a unique 5-bit key $K$ to each permutation $P$ associated with $\Omega$, with $K \in \{0, 1\}^5$ and $|K| = 32$
  5: Keys beyond the first 24 are defined as "ZERO," rendering the output null for any key from 24 to 31
  6: **for** $k = 0$ to $23$ **do**
  7:     Map key $K_k$ to a specific permutation $P_k$ in $\Omega$
  8: **end for**
  9: **Integrate Switch** $\Omega$ **with Router** $R$ **to form** $R_{obf}$**:**
10: **for** each $(i,j)$ in $S$ where $M_{final,ij} = 1$ **do**
11:     Select key $K$ corresponding to the desired permutation $P$ in $\Omega$ for $(i,j)$
12:     Apply $P(K)$ from $\Omega$ to $(i,j)$, updating the connection in $M_{final,ij}$
13: **end for**
14: **return** Obfuscated Router $R_{obf}$ =0

---

The resulting obfuscated router $R_{obf}$ represents the router after integrating with $\Omega$, effectively transforming network topology. By inserting $\Omega$ dynamically into each router to reconfigure the signal paths based on unique keys, POTENT introduces a robust layer of protection, while the updated connectivity matrix $M_{final}^{(R_{obf})}$ captures the obfuscated topology.

The POTENT integration flow, illustrated in Fig. 19, seamlessly incorporates the obfuscation process into the standard SoC design flow. The router $R$ is transformed into its obfuscated counterpart $R_{obf}$, effectively concealing the original network architecture while maintaining operational integrity. Fig. 20 demonstrates the effectiveness of the obfuscation switch $\Omega$ in an NoC-based multi-core design.

Fig. 20(a) presents the original network configuration $S_{orig}$, which is preserved under the correct key via the identity permutation function $P_{id}$.

$$P_{id}(S_{orig}) = S_{orig}$$

In the correct topology, the 9 IPs (IP1 to IP9) are connected with the corresponding router (R1 to R9), respectively, under the key 100, 101, 011 that is applied to the obfuscation switch *I, II, III*. In contrast, alternative keys generate distinct topologies $S'$, altering the NoC topology landscape as depicted in Fig. 20(b), 20(c), and 20(d). In these three topologies, we insert a different key, which leads to a unique topology, though they cannot provide the correct functionality. For example, in Fig. 20(c), we see that the correct key 011 is applied to obfuscation switch *III*, making the following IPs to router connection the correct topology. However, the obfuscation switches *I* and *II* have the incorrect key (011, 001) applied, respectively, which results in the following IP to router connection being different compared to the correct topology. These changes are quantified by a permutation function $P$, where each key $k \in K$ maps $S_{\text{orig}}$ to a new permutation $S'$, representing a reconfigured topology.

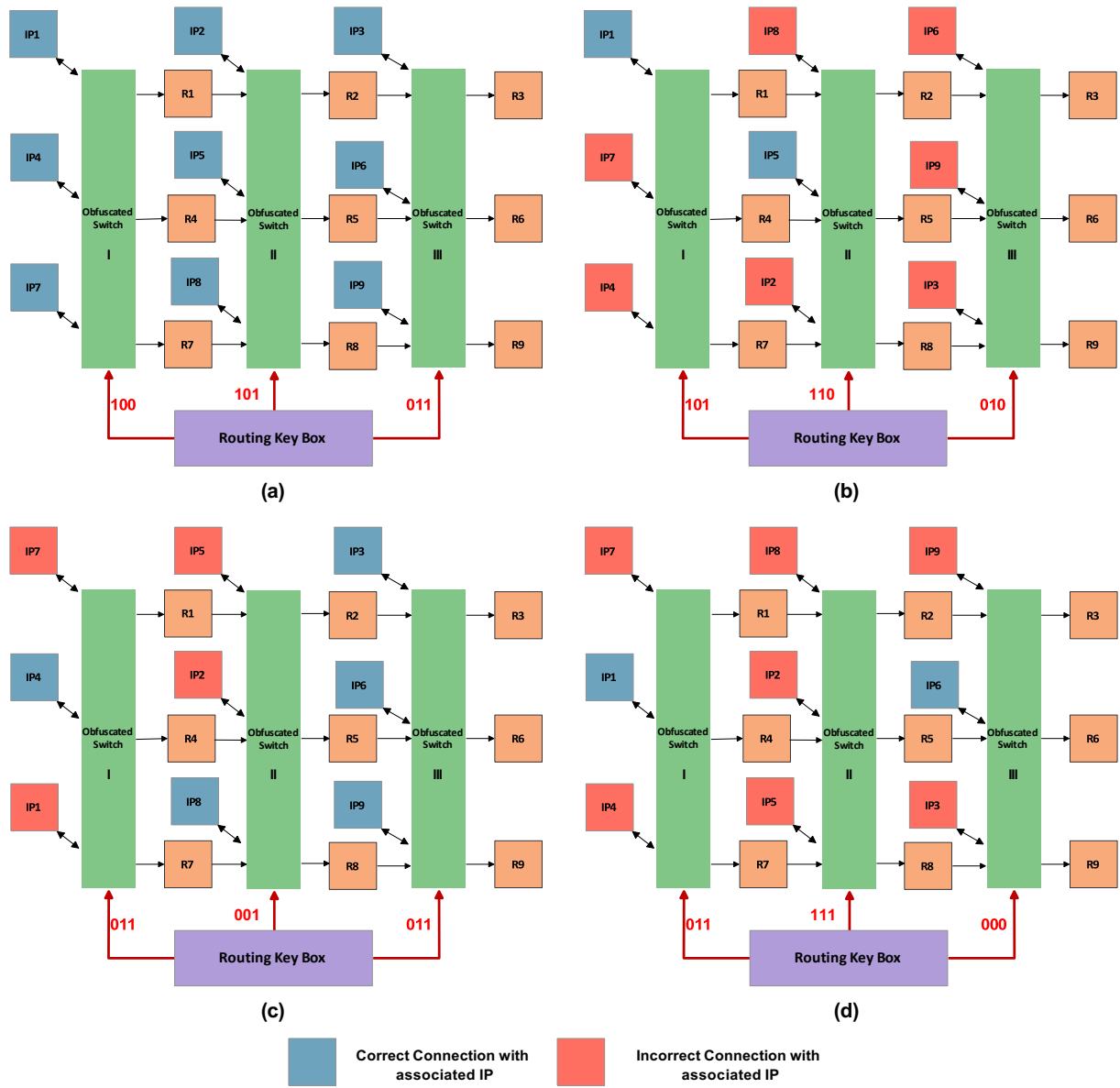

Figure 20: Different Functional Topologies under various Keys

$$P(k)(S_{\text{orig}}) = S', \qquad k \not\models \text{key}_{\text{correct}}$$

This transformation for router-to-IP connections introduces a permutation space that inherently increases the combinatorial diversity and, consequently, the system's security. Each non-identity permutation adds a layer of obfuscation, ensuring that the true network topology remains concealed without the correct key.

## Evaluation and Security Analysis

In the assessment of the POTENT framework, we employed the Quartus Pro 23.2 to construct a sophisticated SoC. This SoC contains a processing unit, memory unit, and communication unit, and all of these IPs were integrated with an NoC fabric. The POTENT framework was applied after generating the initial pre-synthesis netlist from Quartus Platform Designer and subsequent post-synthesis in DesignNavigator, with integrity checks performed by Synopsys Spyglass to ensure the absence of combinational loops. The synthesized unobfuscated SoC comprises a total of approximately $10K$ gates.

## Security Evaluation against SAT Attacks

In our security evaluation against SAT attacks, we have rigorously assessed the resistance of POTENT within an obfuscated SoC context. Traditional SAT-based attacks that target isolated hardware modules are ineffective against POTENT, which secures the NoC on a holistic SoC scale. Advancing beyond the methodologies applied in [19], we have fortified a comprehensive toolchain augmentation, encapsulated by a suite of TCL scripts and open-source utilities, catalyzing the transition from gate-level netlist to an SAT-compatible format. This paradigm shift is symbolized by $F : Net_{post} \rightarrow Net_{SAT}$, where $Net_{post}$ is the gate-level netlist and $Net_{SAT}$ is the netlist formatted for SAT analysis.

Our security evaluation flow on POTENT is shown in Fig. 21. From the formatted netlist $Net_{SAT}$, we get the combinational outputs, which then feed into the SAT attack tool $\Sigma_{SAT}$ to find the recovered key $K_{recovered}$. This $K_{recovered}$ is then compared with the correct key $K_{correct}$. We have found that the $K_{recovered}$ obtained from the SAT attack was incorrect, *i.e.*, different from $K_{correct}$. In fact, $K_{recovered}$ corresponds to a different topology $T_{alt}$. As POTENT is designed in this way to generate each different topology under each unique key, this key-topology binding is expressed by $\Phi(K_{recovered}) \not\models T_{original}$. The implications are twofold.

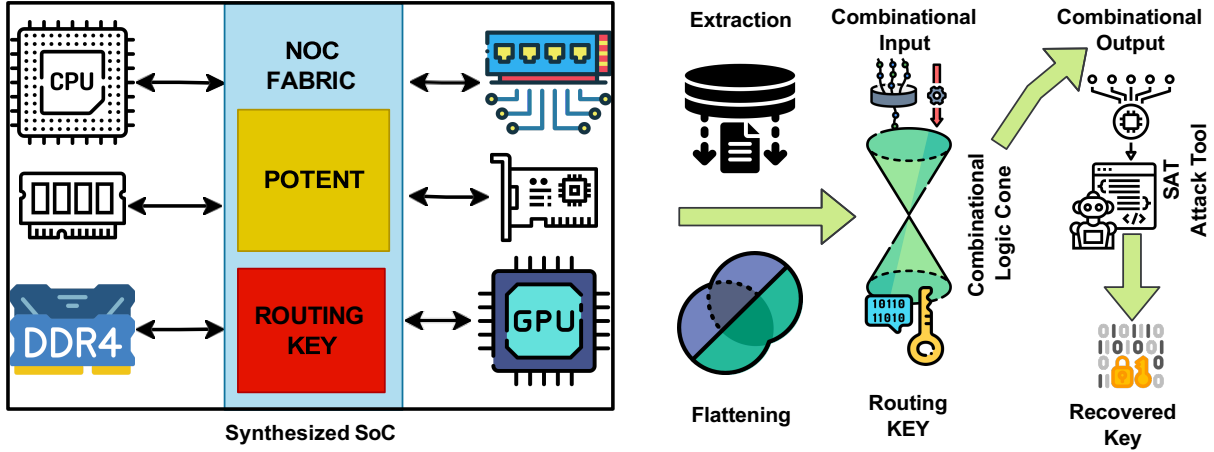

Figure 21: SAT Attack Flow

1. The $K_{recovered}$ that did not reinstate the original functionality underscores POTENT's unique construction, where $\Phi : K \overset{1}{\rightarrow} T$ is a bijection for only the correct key, $K_{correct}$, ensuring that any $K \not\models K_{correct}$ leads to an altered topology $T_{alt}$.

2. From the attacker's perspective, the SAT attack does not recover the original topology $T_{original}$ and, hence, the original functionality of the design.

In summary, the SAT attack's inability to compromise the original design via $K_{recovered}$ affirms the efficacy of POTENT.

## Resource Overhead Analysis

The comparative resource overhead analysis has been shown in Fig. 22. We used gate count, power consumption, area, and delay as the primary evaluation metrics to quantify the resource overhead between the unobfuscated and obfuscated design. Compared to unobfuscated design, design with POTENT has an overhead of 8.32 % in gate count. This trend persists in the area overhead, which stands at 8.47 %. The integration of POTENT also exerts on the total power consumption, resulting in a 9.83% increase. The NoC fabric in the original unobfuscated SoC is responsible for approximately 25 % of the total gate count, and the SoC itself is comparatively small. If the same obfuscation is applied to an industrial SoC where NoC fabric typically accounts for a much smaller gate count (the footprint is dominated by large IPs like the CPU), the overhead would obviously be much less. Considering this fact, the resource overhead on area, power, and gate count, which is between 8 % to 10 % in this SoC implementation, is not that significant. On the other hand, as the primary communication backbone, this NoC fabric inherently exhibits the highest delay, which is 81.58 % within the SoC. However, the obfuscation process introduces only 2 % overhead in the final obfuscated design. Overall, POTENT does not incur significant resource overhead, considering the security robustness it provides.

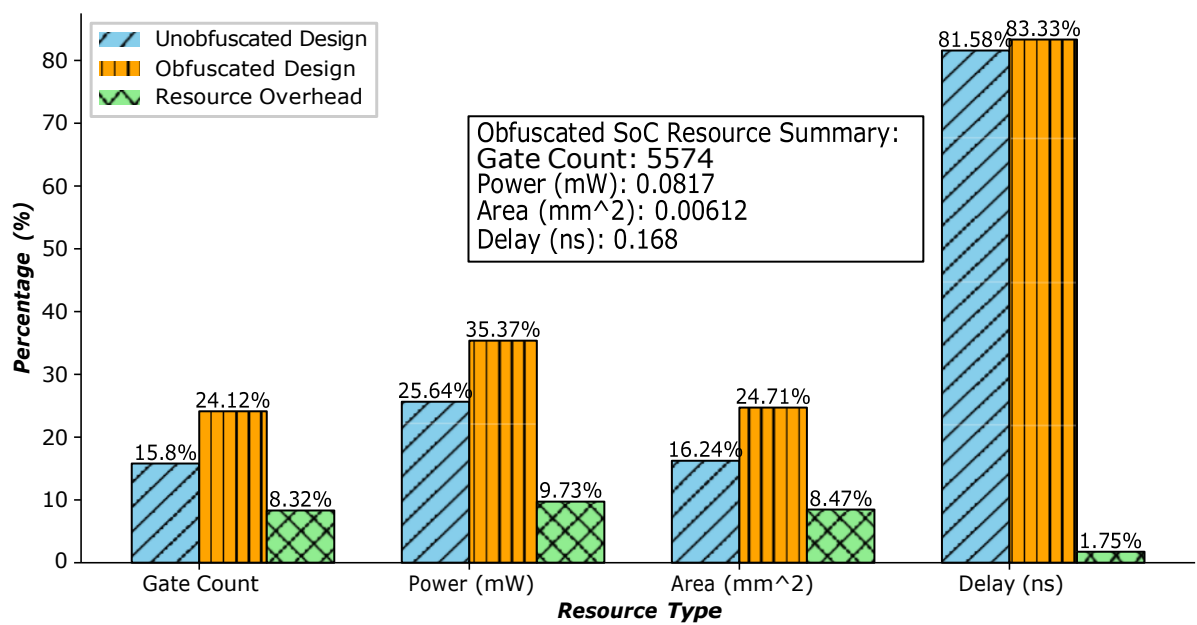

Figure 22: Resource Utilization Analysis

# 4 Research Milestones: Extending SoC Interconnect Security to Chiplets

The evolution of semiconductor technology has led to a significant shift from traditional monolithic System-on-Chip (SoC) designs to chiplet-based architectures. Chiplets are modular functional units, each designed to perform specific tasks, which are integrated on a single package through high-performance interconnects. This paradigm shift addresses the limitations of Moore's Law and the increasing challenges in yield, cost, and design complexity faced by monolithic SoCs. By decoupling functionalities into smaller dies, chiplets offer scalability, flexibility, and cost-efficiency. For instance, they allow the use of advanced process nodes for performance-critical components while leveraging mature, cost-effective nodes for others.

Chiplet technology is revolutionizing modern microelectronics by enabling heterogeneous integration. Diverse technologies such as GPUs, DSPs, memory, and custom accelerators can now coexist on the same package, optimized for specific applications. This heterogeneity enhances system performance while reducing development costs and improving manufacturing yields. Moreover, chiplet-based designs enable better customization and faster time-to-market by reusing pre-validated modules. In Fig 26, we have shown a chiplet based SoC. There are two types of communication in a chiplet based system which are inter-chiplet communication and intra-chiplet communication.

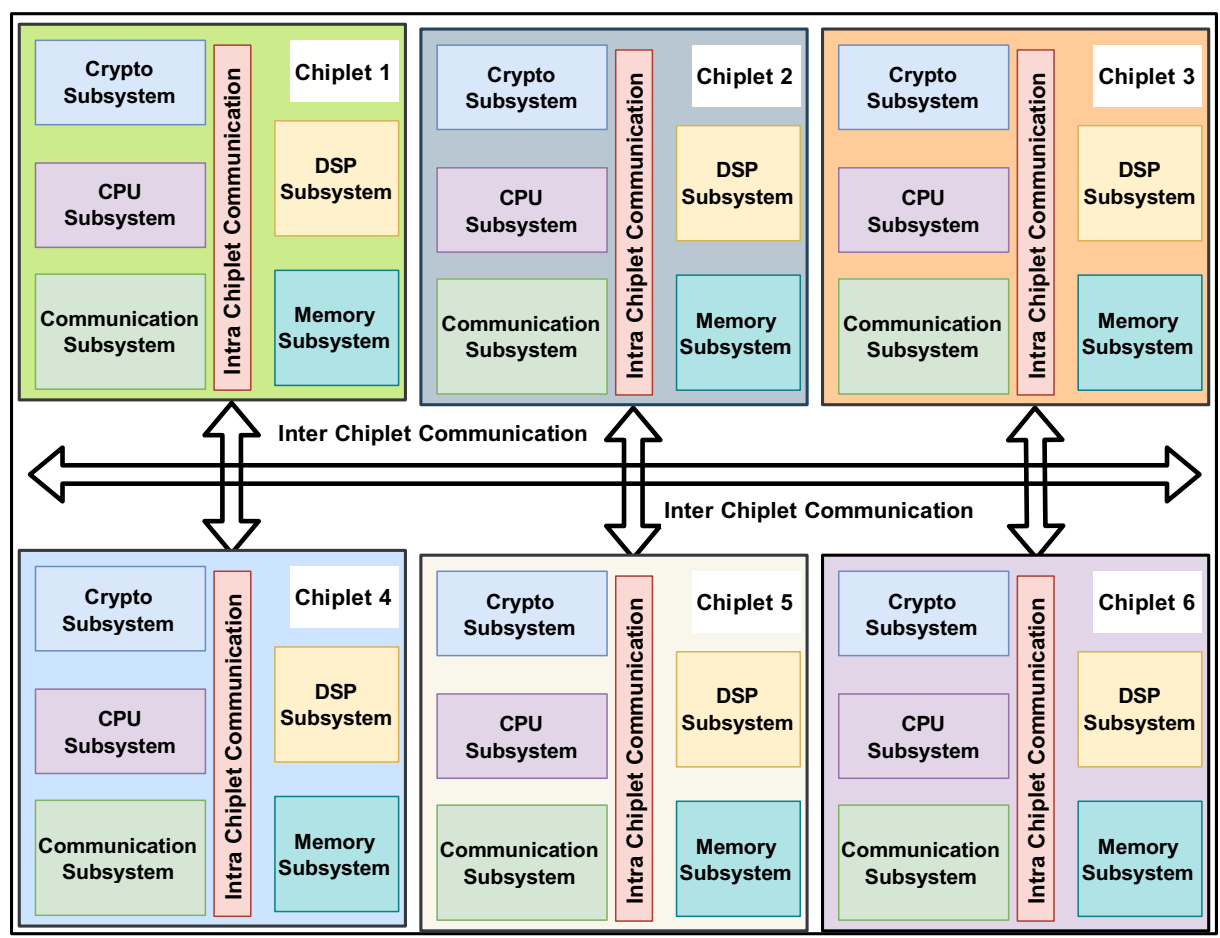

Figure 23: Chiplet based modern microelectronic system

## Inter-Chiplet Communication

Inter-chiplet communication involves data transfer between different chiplets within the same package. This type of communication relies on high-bandwidth, low-latency interconnects such as Advanced Interface Bus (AIB) or Universal Chiplet Interconnect Express (UCIe). Securing inter-chiplet communication is critical, as vulnerabilities in these links can expose sensitive data, enable unauthorized access, or disrupt system functionality. The challenge lies in ensuring both performance efficiency and robust security, particularly in heterogeneous designs where chiplets may originate from multiple vendors.

## Intra-Chiplet Communication

Intra-chiplet communication refers to the data exchange within an individual chiplet, typically between functional units like processors, memory controllers, or accelerators. While intra-chiplet communication often benefits from well-established SoC design practices, ensuring security remains vital. Compromised intra-chiplet communication can lead to breaches in internal operations or act as an entry point for attackers to compromise the entire system. Techniques such as secure Network-on-Chip (NoC) designs and hardware-based encryption can help mitigate these risks.

However, this shift introduces unique challenges, particularly in the design and security of interconnects that link these chiplets. The communication infrastructure between chiplets becomes a critical vulnerability, especially when components from different vendors are integrated. Securing these interconnects is imperative to protect sensitive data, prevent reverse engineering, and ensure the overall integrity of the system. As the semiconductor industry rapidly adopts chiplet-based designs—demonstrated by advancements from companies like AMD and Intel—the need for robust, secure interconnect solutions has become paramount. This proposal aims to extend prior work on SoC interconnect security to address these emerging challenges in chiplet-based systems.

## Security Challenges in Chiplet-Based Interconnects

The transition from traditional monolithic System-on-Chip (SoC) designs to chiplet-based architectures introduces significant advantages in scalability, cost reduction, and heterogeneity. However, this shift also creates new security challenges due to the distributed nature of chiplet-based interconnects and the involvement of diverse, often untrusted entities in the design and manufacturing pipeline. This section outlines the key security challenges in chiplet-based interconnects, emphasizing the need for robust mechanisms to ensure secure and reliable communication.

### Distributed Design

Unlike SoCs, chiplet-based systems rely on inter-chiplet communication across multiple physical dies within a single package. These interconnects must meet the demanding requirements of high-bandwidth and low-latency communication, often leveraging standards like Universal Chiplet Interconnect Express (UCIe) and Compute Express Link (CXL). However, the distributed nature of these designs creates vulnerabilities at the interfaces between chiplets, where adversaries can exploit weak points to disrupt communication or exfiltrate sensitive data.

### Untrusted Entities

Chiplet-based architectures depend on a fragmented supply chain involving third-party IP vendors, untrusted foundries, interposer manufacturers, and assembly facilities. This reliance increases the attack surface, as malicious actors within the supply chain can engage in reverse engineering, tampering, Trojan insertion, or overproduction of unauthorized chiplets. The heterogeneity of chiplets, sourced from multiple vendors, further complicates the security landscape by introducing varying trust levels across the system.

### Intra-Chiplet Security

Within individual chiplets, securing the internal communication fabric, such as the Network-on-Chip (NoC), is critical. The NoC connects multiple functional units within the chiplet, including processors, memory, and accelerators. Attackers targeting intra-chiplet communication can reverse-engineer NoC topologies or exploit vulnerabilities in routing logic, leading to the exposure of sensitive data or denial-of-service attacks. Techniques such as topology obfuscation and encrypted communication must be employed to safeguard intra-chiplet NoC designs.

## Inter-Chiplet Security

Inter-chiplet communication, facilitated by interconnects like UCIe or CXL, is another major security concern. Adversaries can target these interconnects to intercept or manipulate data during transfer. Additionally, ensuring the authenticity of chiplets and preventing unauthorized access or data leakage is vital. Authentication mechanisms and secure handshaking protocols must be implemented to ensure that only trusted chiplets participate in the system.

## Key Threats

Several significant threats loom over chiplet-based interconnects, including Reverse Engineering, Tampering and Trojan Insertion, Physical Interface Exploitation. We are going to explore three significant security concerns in Chiplet based design.

## Secure Inter-Chiplet Communication in Compromised Chiplet-Based Systems

As chiplet-based architectures emerge as the next frontier in microelectronics, their distributed and modular nature introduces unprecedented security challenges. Unlike monolithic System-on-Chip (SoC) designs, chiplet-based systems rely on inter-chiplet communication across multiple dies, often sourced from diverse vendors. While this modularity enhances scalability, functionality, and cost-efficiency, it also opens doors to vulnerabilities that can compromise the security and integrity of the entire system. Securing inter-chiplet communication, particularly in the presence of untrusted or compromised chiplets, becomes a cornerstone of ensuring the trustworthiness of these systems.

**Distributed Design:** In chiplet-based systems, inter-chiplet communication traverses physical interfaces, such as interposers or advanced packaging technologies, introducing risks of data interception, protocol manipulation, and physical tampering. The distributed nature means that security breaches in one chiplet or communication channel could compromise the entire system.

**Untrusted Entities:** The reliance on a supply chain involving multiple vendors—including untrusted or partially trusted manufacturers and assembly houses—exacerbates the security risks. A compromised chiplet could inject malicious payloads, leak sensitive data, or disrupt communication protocols. Additionally, attackers can exploit weaknesses in the interconnect interfaces, such as Universal Chiplet Interconnect Express (UCIe) or Compute Express Link (CXL), to compromise the system.

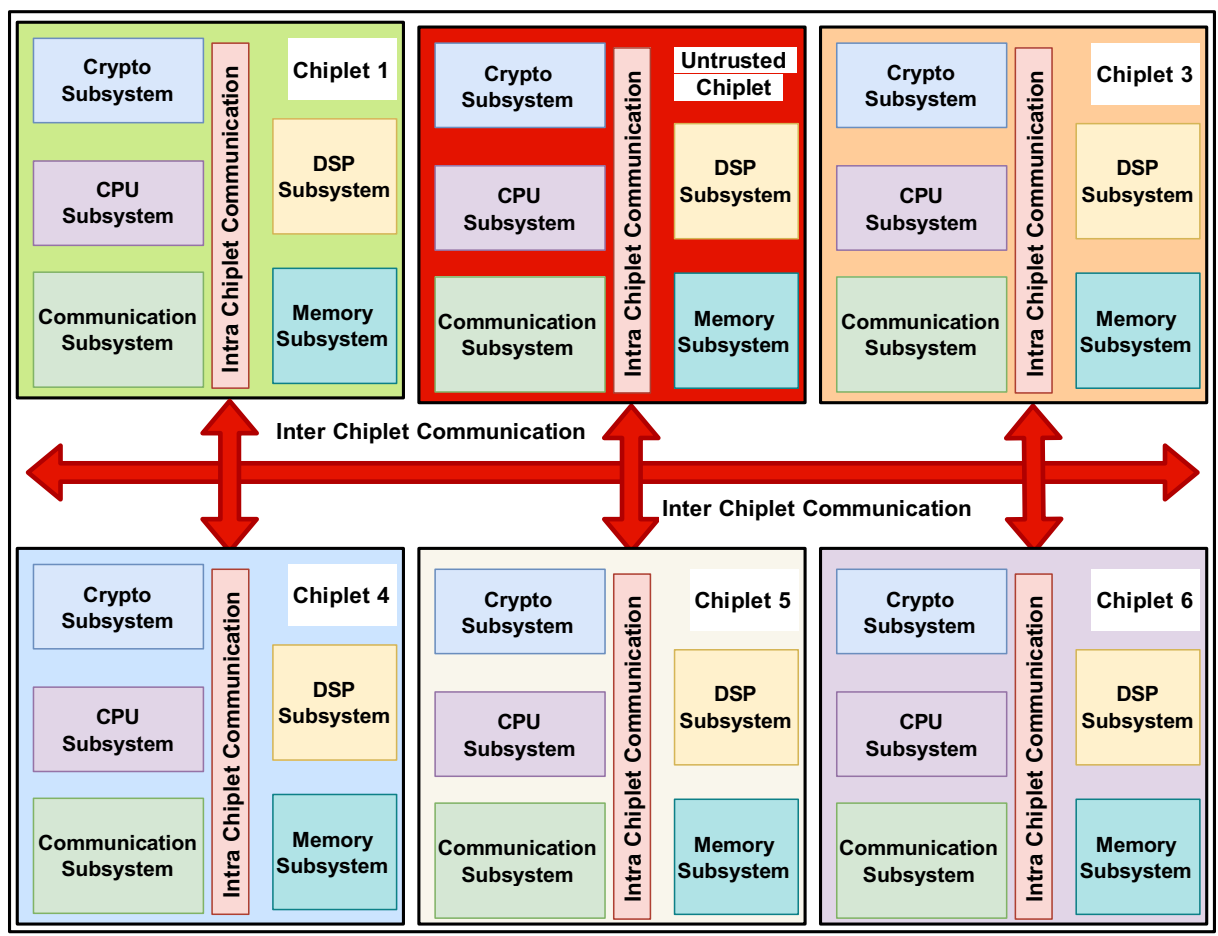

Figure 24: Interconnect security scenario in compromised chiplet based design

**Authentication Challenges:** Unlike monolithic SoCs, where trust can often be established at the design level, chiplet-based systems require robust authentication mechanisms to validate the legitimacy and behavior of each chiplet within the system. This process must occur dynamically, as the chiplets interact during runtime.

## Objectives of the Research

To address these security challenges, this research focuses on designing mechanisms and frameworks that ensure secure communication between chiplets, even in scenarios where some chiplets may be compromised. The objectives include:

**Comprehensive Vulnerability Analysis:** Investigate existing inter-chiplet communication protocols and architectures to identify potential weaknesses that can be exploited by adversaries.

**Development of Resilient Security Mechanisms:** Propose novel security techniques to authenticate chiplets, secure data transmission, and protect inter-chiplet communication against tampering, reverse engineering, and physical attacks. Framework for Compromised Scenarios: Develop a framework that can detect compromised chiplets and isolate them, ensuring that the rest of the system continues to operate securely without significant performance degradation.

**Research Questions**

This study is guided by the following key questions:

- What vulnerabilities exist in inter-chiplet communication protocols? Understanding the weaknesses is critical to designing effective countermeasures.

- How can we identify compromised chiplets in a system? This includes developing dynamic runtime mechanisms that can detect malicious behaviors without reliance on prior knowledge of specific attacks.

- What enhancements or modifications to communication protocols can mitigate security risks? Exploring changes to interconnect standards like UCIe and CXL that maintain compatibility while improving resilience.

# Protecting Inter-Chiplet Interconnect Design from Reverse Engineering

While the modular design of chiplet systems enhances scalability and performance, it also presents significant security challenges, particularly from reverse engineering attacks. Reverse engineering of inter-chiplet interconnects can compromise proprietary protocols, expose sensitive data flows, and enable unauthorized cloning or manipulation of the design. This section addresses the critical need for securing inter-chiplet interconnects against such threats.

## Reverse Engineering Risks

Reverse engineering allows adversaries to reconstruct the design and functionality of inter-chiplet interconnects. This involves analyzing the physical implementation, extracting communication protocols, and deducing system-level interactions. Such attacks may lead to:

**Intellectual Property (IP) Theft:** Attackers can replicate proprietary communication protocols, undermining the competitive advantage of the design.

**System Integrity Breaches:** Exposing the data flow between chiplets increases the risk of malicious interventions, including data tampering or denial-of-service attacks.

**Trojan Insertion:** Adversaries may implant hardware Trojans in the interconnect layer, enabling covert data leakage or system failures during runtime.

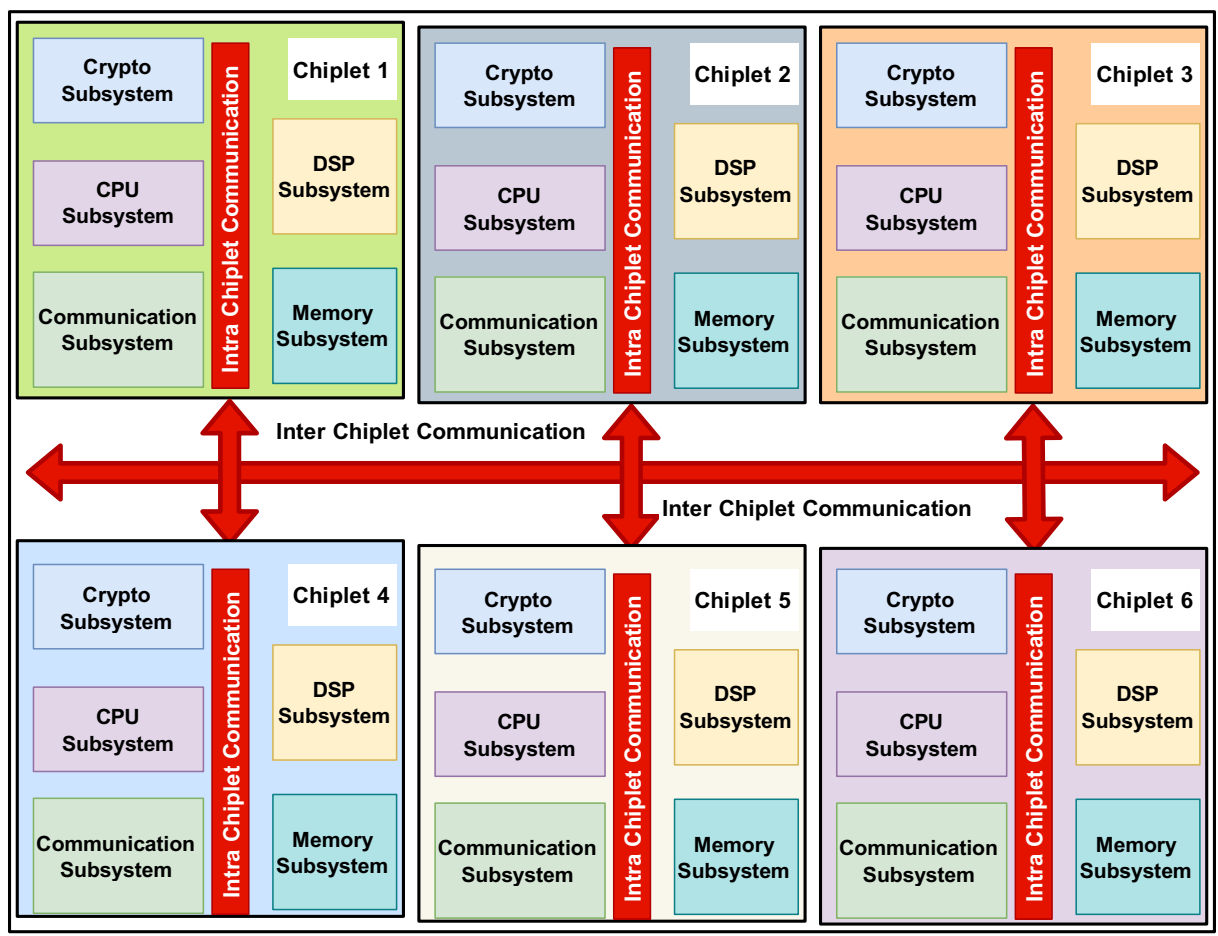

Figure 25: Interconnect security scenario in Reverse Engineering attacks

## Research Objectives

- Develop robust obfuscation techniques tailored for inter-chiplet interconnects.

- Evaluate the trade-offs between security and performance in protected interconnect designs.

- Establish a framework for real-time anomaly detection to identify and mitigate reverse engineering attempts.

## Secure Inter-Chiplet Communication Protocols Against Unauthorized Access

Inter-chiplet communication traverses shared interconnects, creating vulnerabilities that adversaries can exploit to access sensitive data or manipulate system behavior. Addressing these challenges requires the design and implementation of robust protocols that provide confidentiality, integrity, and authentication.

**Authentication of Chiplets:** Unverified chiplets from untrusted vendors pose a significant security risk. An adversary could introduce malicious chiplets that leak information or interfere with legitimate communication.

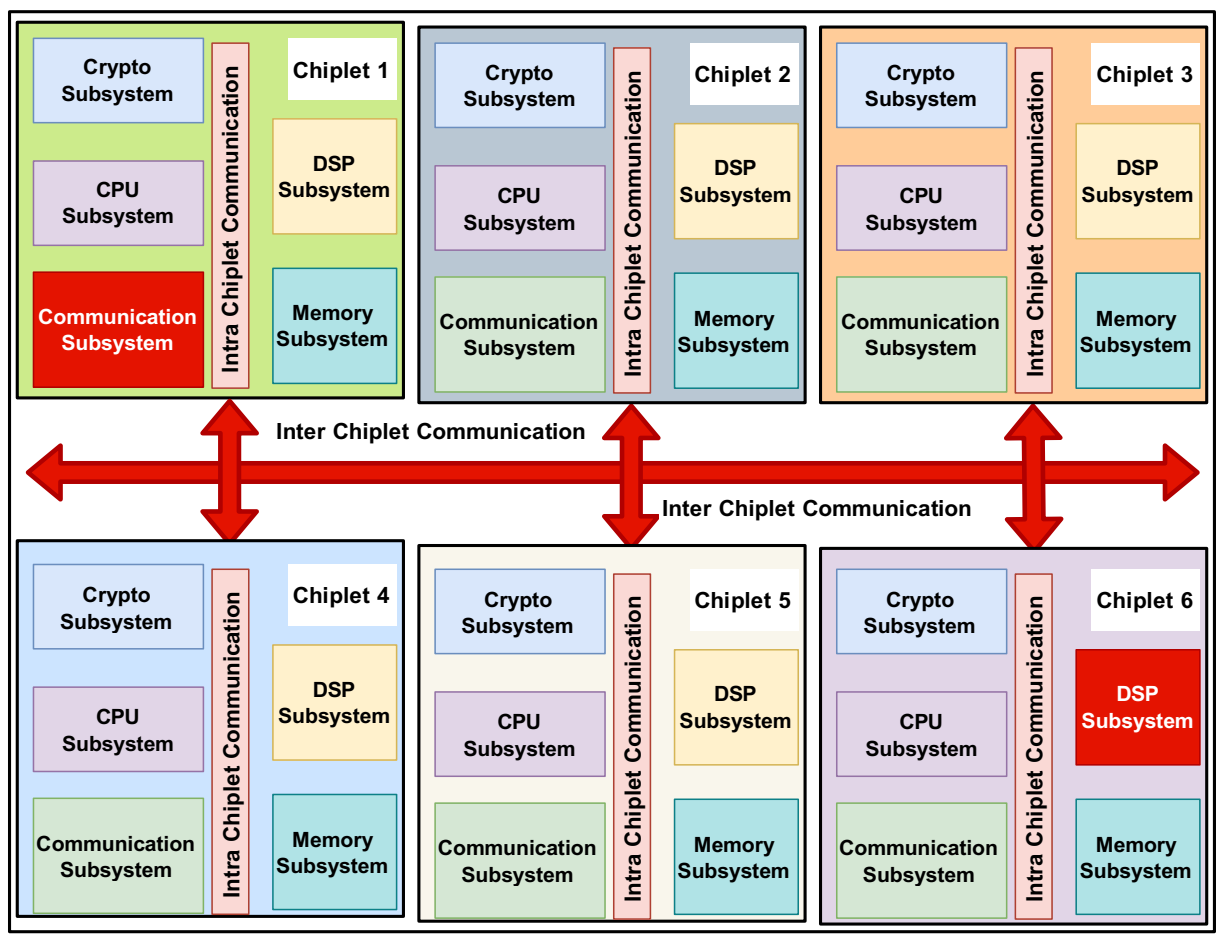

Figure 26: Interconnect security scenario in unauthorized protocol accessing

**Data Confidentiality and Integrity:** Inter-chiplet communication must ensure that transmitted data is encrypted and protected from tampering during transit. Without this, sensitive information is exposed to interception or corruption.

**Protocol Robustness:** Existing interconnect standards, such as UCIe and CXL, lack inherent mechanisms to authenticate chiplets or prevent unauthorized communication. These limitations necessitate enhancements to secure the overall architecture.

## Research Questions

The evolution of chiplet-based architectures brings forth significant security concerns, particularly in inter-chiplet communication protocols. To address these challenges, the following research questions guide the exploration of secure inter-chiplet communication:

- How can authentication be enforced between chiplets to prevent unauthorized access?

- What cryptographic techniques can provide confidentiality and integrity for inter-chiplet data communication

- How can existing interconnect standards be extended to include security primitives?

- What metrics can evaluate the effectiveness of secure inter-chiplet protocols?

- How can compromised or malicious chiplets be detected and isolated in real time?

- What are the trade-offs between security, performance, and scalability in inter-chiplet communication protocols?

# 5    Conclusion

In this proposal, we address the critical security challenges associated with modern System-on-Chip (SoC) and emerging chiplet-based architectures. As the semiconductor industry transitions from monolithic designs to distributed chiplet-based systems, ensuring the security and reliability of interconnects becomes paramount. Through our prior works, ObNoCs and POTENT, we have demonstrated robust methodologies to safeguard NoC fabrics against reverse engineering and synthesis-induced vulnerabilities. Building on this foundation, this research extends the focus to inter- and intra-chiplet communication, proposing novel obfuscation techniques, secure interconnect protocols, and authentication mechanisms to protect against unauthorized access, data leaks, and malicious modifications.

The proposal outlines a comprehensive plan to investigate and mitigate vulnerabilities in chiplet-based interconnects, leveraging both theoretical frameworks and practical implementations. We aim to bridge the gap between traditional SoC security paradigms and the unique challenges of distributed chiplet systems. By addressing key security issues such as reverse engineering, Trojan insertion, and compromised communication channels, this research seeks to pave the way for a secure and resilient semiconductor ecosystem.

The proposed research contributes to the field of hardware security by developing scalable solutions for protecting interconnects in both SoC and chiplet-based architectures. With a clearly defined timeline, actionable research milestones, and a focus on practical implementation, this work aims to establish a secure foundation for next-generation microelectronics. Ultimately, this research will enable the deployment of trusted, high-performance systems for critical applications in domains such as defense, AI accelerators, and advanced computing.

# 6    Tentative Timeline

The research presented in this proposal is structured with a clear timeline to ensure the systematic progression of objectives, deliverables, and impactful contributions to the field of secure interconnects. Below is the detailed timeline outlining key milestones and research phases:

- Fall 2023: Successfully developed ObNoCs (Obfuscated Network-on-Chip) methodology, addressing reverse engineering attacks on NoC fabric. Published this work in the ACM Transactions on Embedded Computing Systems (TECS) journal, marking the first work to tackle such security challenges comprehensively.

- Fall 2024: Designed and implemented PoTeNt (Post-Synthesis Obfuscation Techniques), focusing on resolving optimization challenges induced by synthesis tools. Presented this contribution at the IEEE System-on-Chip (SOCC) conference

- Spring 2025: Currently preparing a journal article on PoTeNt, further elaborating on its findings and technical contributions. Initiated research on interconnect vulnerabilities within chiplet-based systems, focusing on both inter-chiplet and intra-chiplet communication security. Targeting the submission of a conference paper highlighting identified vulnerabilities and proposed solutions for chiplet-based interconnects.

- Summer 2025: Planning to publish a journal article detailing a secure interconnect framework for compromised chiplet-based designs. Begin exploration and development of techniques to counter unauthorized protocol access in inter-chiplet communication, contributing to the broader scope of secure communication frameworks.

- Fall 2025: Aim to publish a comprehensive journal article addressing a unified interconnect security solution for modern microelectronics. This work will integrate learnings from both SoC and chiplet-based architectures, providing a holistic security framework applicable to next-generation systems.

This structured timeline highlights the alignment of research efforts with evolving technological challenges, emphasizing the incremental yet impactful contributions aimed at advancing interconnect security across modern microelectronics.